\documentclass[fleqn,usenatbib]{mnras}

\usepackage{newtxtext,newtxmath}

\usepackage[T1]{fontenc}

\DeclareRobustCommand{\VAN}[3]{#2}
\let\VANthebibliography\thebibliography
\def\thebibliography{\DeclareRobustCommand{\VAN}[3]{##3}\VANthebibliography}

\usepackage{graphicx}	
\usepackage{amsmath}	

\title[Subhalo inference with streams]{Inferring dark matter subhalo properties from simulated subhalo-stream encounters}

\author[T. Hilmi et al.]{
\parbox{\textwidth}{
\Large
Tariq~Hilmi,$^{1}$
Denis~Erkal,$^{1}$
Sergey~E.~Koposov,$^{2,3,4}$
Ting~S.~Li,$^{5,6,7}$
Sophia~Lilleengen,$^{8}$
Alexander~P.~Ji,$^{9,10}$
Geraint~F.~Lewis,$^{11}$
Nora~Shipp,$^{12}$
Andrew~B.~Pace,$^{12}$
Daniel~B.~Zucker,$^{13,14}$
Guilherme~Limberg,$^{15}$
and Sam~A.~Usman$^{9,10}$
\begin{center} ($S^5$ Collaboration) \end{center}
}
\vspace{0.4cm}
\\
\parbox{\textwidth}{
$^{1}$ Department of Physics, University of Surrey, Guildford GU2 7XH, UK\\
$^{2}$ Institute of Astronomy, University of Cambridge, Madingley Road, Cambridge CB3 0HA, UK\\
$^{3}$ Institute for Astronomy, University of Edinburgh, Royal Observatory, Blackford Hill, Edinburgh EH9 3HJ, UK\\
$^{4}$ Kavli Institute for Cosmology, University of Cambridge, Madingley Road, Cambridge CB3 0HA, UK\\
$^{5}$ Department of Astronomy \& Astrophysics, University of Toronto, 50 St. George Street, Toronto ON, M5S 3H4, Canada\\
$^{6}$ Dunlap Institute for Astronomy \& Astrophysics, University of Toronto, 50 St George Street, Toronto, ON M5S 3H4, Canada\\
$^{7}$ Data Sciences Institute, University of Toronto, 17th Floor, Ontario Power Building, 700 University Ave, Toronto, ON M5G 1Z5, Canada\\
$^{8}$ Institute for Computational Cosmology, Department of Physics, Durham University, South Road, Durham DH1 3LE, UK\\
$^{9}$ Department of Astronomy \& Astrophysics, University of Chicago, 5640 S Ellis Avenue, Chicago, IL 60637, USA\\
$^{10}$ Kavli Institute for Cosmological Physics, University of Chicago, Chicago, IL 60637, USA\\
$^{11}$ Sydney Institute for Astronomy, School of Physics, A28, The University of Sydney, NSW 2006, Australia\\
$^{12}$ McWilliams Center for Cosmology, Carnegie Mellon University, 5000 Forbes Ave, Pittsburgh, PA 15213, USA \\
$^{13}$ School of Mathematical and Physical Sciences, Macquarie University, Sydney, NSW 2109, Australia \\
$^{14}$ Macquarie University Research Centre for Astrophysics and Space Technologies, Sydney, NSW 2109, Australia \\
$^{15}$ Universidade de S\~ao Paulo, IAG, Departamento de Astronomia, SP 05508-090, S\~ao Paulo, Brazil\\
}
}

\date{Accepted XXX. Received YYY; in original form ZZZ}

\pubyear{2024}

\begin{document}
\label{firstpage}
\pagerange{\pageref{firstpage}--\pageref{lastpage}}
\maketitle

\begin{abstract}
In the cold dark matter paradigm, our Galaxy is predicted to contain >10000 dark matter subhaloes in the $\mathrm{10^5\,-10^8\,M_\odot}$ range which should be completely devoid of stars. Stellar streams are sensitive to the presence of these subhaloes, which can create small-scale features in streams if they pass closely enough. Modelling these encounters can therefore, potentially recover the subhalo’s properties. In this work, we demonstrate this for streams generated in numerical simulations, modelled on eccentric orbits in a realistic Milky Way potential, which includes the Large Magellanic Cloud and the subhalo itself. We focus on a mock model of the ATLAS-Aliqa Uma stream and inject a $10^7 \, {\rm M}_\odot$ subhalo, creating a similar discontinuous morphology to current observations. We then explore how well subhalo properties are recovered using mock stream observations, consisting of no observational errors, as well as assuming realistic observational setups. These setups include present day style observations, and what will be possible with 4MOST and Gaia DR5 in the future. We show that we can recover all parameters describing the impact even with uncertainties matching existing data, including subhalo positions, velocities, mass and scale radius. Modelling the subhalo on an orbit instead of assuming an impulse approximation, we greatly reduce the degeneracy between subhalo mass and velocity seen in previous works. However, we find a slight bias in the subhalo mass ($\sim 0.1$ dex). This demonstrates that we should be able to reliably extract the properties of subhaloes with stellar streams in the near future.

\end{abstract}




\section{Introduction}

In the cold dark matter paradigm, structure formation proceeds in a hierarchical fashion \citep[e.g.][]{1978MNRAS.183..341W}. In the early Universe, small-scale overdensities collapse first, forming small haloes that merge together to form larger and larger haloes. As a result, the dark matter halo of a galaxy like the Milky Way is expected to contain a plethora of dark matter subhaloes spanning a wide range of masses at the present day \citep{klypin1999,moore99,springel08}. While the most massive of these dark matter subhaloes host dwarf galaxies, below a mass of $\sim10^8 \, {\rm M}_\odot$ they are believed to be devoid of stars \citep[e.g.][]{Somerville_etal_2015,Jethwa_etal_2018,Nadler_etal_2021}.

Detecting these starless dark matter subhaloes would be a confirmation of the dark matter paradigm and would allow us to probe the properties of the dark matter particle itself. For example, in warm dark matter, these subhaloes should only exist down to the free-streaming scale below which the thermal motion of particles in the early Universe washes out substructure \citep{Bond_Szalay_1983,Bode_etal_2001}. This free-streaming scale is set by the mass of the warm dark matter particle and thus the detection of low-mass subhaloes can constrain the mass of the dark matter particle. Similarly in fuzzy dark matter, low mass subhaloes are suppressed due to quantum pressure \citep[e.g.][]{Hui_etal_2017}. Thus these dark matter subhaloes act as a magnifying glass with which we can examine the properties of the dark matter particle.

Stellar streams in the Milky Way have long been thought of as an excellent detector of these dark matter subhaloes \citep[e.g.][]{ibata02,johnston02,siegal08,carlberg09}. These streams form as globular clusters (GCs) or dwarf galaxies tidally disrupt in the presence of the Milky Way. Once stripped, the stars in the stream follow roughly the same orbit and thus the streams appear to roughly trace great circles on the sky. Despite their seeming coherence, the streams are fragile structures. The passage of a nearby subhalo imparts velocity kicks on the stream stars which subsequently changes their orbit around the Milky Way, creating gaps and wiggles in the stream \citep[e.g.][]{yoon11,Carlberg_2012,erkal15a}.

In order to extract the properties of subhaloes with stellar streams, two different approaches have been proposed. \cite{Bovy_etal_2017} showed that the statistical properties of the perturbed stream (e.g. the power spectrum of the density fluctuations) can be used to estimate the population of subhaloes that the stream has been exposed to. This is done by repeatedly forward modeling the stream in the presence of varying amounts of substructure and comparing the statistical property of the resulting streams with the observed stream. \cite{Banik_etal_2021a,Banik_etal_2021b} used this approach to set one of the tightest constraints on the mass of the dark matter particle using observations of the GD-1 \citep{GD1_disc} and Palomar 5 \citep{Pal5_disc} streams. While this approach can be used to measure the statistical properties of the dark matter subhalo population, it cannot be used to identify individual subhalo encounters.

The second approach is to fit individual subhalo impacts and extract each impactor's properties. \cite{erkal15b} use an analytic model for a stream initially on a circular orbit that is perturbed by a subhalo using the impulse approximation. They showed that it is possible to extract all of the properties describing the subhalo's flyby (i.e. the subhalo mass, size, impact position and velocity, and impact time) using observations of the perturbed stream. They further showed that this inference is possible even with realistic observational errors similar to what is possible at the present day. \cite{bonaca19} attempted a fit to the perturbed GD-1 stream first observed in \cite{pricewhelan18}. In order to fit the perturbed GD-1 stream, \cite{bonaca19} assumed that all of the stars in the stream were on the same orbit, i.e. that GD-1 was an infinitely cold stream. They found that perturbers in the mass range $10^6-10^8 {\rm M}_\odot$ could produce similar features to what is observed.

The goal of this work is to build on the work of \cite{erkal15b} and \cite{bonaca19} to fit realistic stream models to perturbed stellar streams. In order to make the stream models as realistic as possible, the models will be generated with the modified Lagrange Cloud stripping technique \citep{gibbons14}. For further realism, we also do not assume the subhalo's perturbation can be described with the impulse approximation. Instead, we inject a subhalo that orbits in the Milky Way with the stream in order to self-consistently perturb it. 

In order to test how well this inference works, we use a mock stream inspired by the ATLAS-Aliqa Uma (AAU) stream. Initially, the two segments of this stream were discovered individually as the ATLAS stream by \citet{koposov14} in the VST ATLAS Survey DR1 \citep{Shanks_etal_2015}, and a separate Aliqa Uma stream by from the Dark Energy Survey from \citet{shipp18}. As a result of their different distances of $\sim$20$\mathrm{\,kpc}$ and $\sim$28$\mathrm{\,kpc}$ respectively, their differing on-sky alignment, and their being $\sim$1$\mathrm{\,\deg}$ apart on the sky, ATLAS and Aliqa Uma were initially thought of as two separate streams. Spectroscopic follow-up by the Southern Stellar Stream Spectroscopic Survey \citep[$S^5$;][]{li19,S5_DR1,li22} then revealed that ATLAS and Aliqa Uma had kinematics and metallicities consistent with a single stream \citep{li21}. \cite{li21} argued they were one stream (dubbed the AAU stream) with a significant `kink'-like perturbation. What was most unusual was that outside the kink itself, the stream was relatively unperturbed, with an otherwise thin, smooth track on either side of the kink. Furthermore, they showed that this highly localized perturbation was difficult to recreate using baryonic structures (e.g. dwarf galaxies impacts, the bar, spiral arms, or giant molecular clouds), hence making it an interesting stream to study for the purpose of probing dark matter substructure. 

Our paper is structured as follows; in Section \ref{sec:code} we describe how we set up a stream-subhalo encounter, including summarising our code and how encounters are described using 8 key parameters. In Section \ref{sec:recoverystuff} we describe our mock impact and outline the different observational uncertainties we use, alongside how we fit the subhalo properties itself. We then outline and describe the results in Section \ref{sec:MCMCresults}, and discuss the implications of these results in Section \ref{sec:discussion}. Finally, we present our conclusions in Section \ref{sec:conclusions}.

\section{An overview of the setup} \label{sec:code}


In this section, we outline the process of generating stellar streams that have been perturbed by a subhalo-like object. We split this into 3 subsections, corresponding to the major steps involved in doing so, with a rough outline detailed below.

Our work aims to fit a perturbed stellar stream in order to extract the properties of the perturber. To do this, we use a setup that follows a consistent set of impact parameters that define when an encounter took place, and the subhalo properties and dynamics relative to the stream. Our procedure of generating a perturbed stream consists of three steps:\\

$\bullet$ We initially generate a stream in the absence of any subhalo. The stream is integrated up until the impact time so that the subhalo's position at the impact time can be determined. \\

$\bullet$ We then place a subhalo at the final snapshot of the first step (i.e. at the impact time), with the impact parameters allowing us to determine its position and velocity at this time. We then integrate the subhalo forward to the present day to get its present-day phase space coordinates. \\

$\bullet$ Finally, we rewind the progenitor and subhalo backward from the present day and then integrate the stream and subhalo up until the present day to generate a stream perturbed by a subhalo. We stress that during the rewinding, our subhalo can affect the progenitor's past orbit, and thus the full effect of the subhalo is self-consistent. While our impact parameters ensure we have that specific encounter, we make no assumption that this is the only impact taking place. The inclusion of the subhalo throughout integration from stream generation to the present day adds the possibility for additional interactions. 

We note our approach assumes that the stream progenitor initially disrupts around the Milky Way producing a stream that is then perturbed by a subhalo. While this is quite a general tool for modelling perturbers, it cannot be used to model a globular cluster stream that forms by first disrupting around a dwarf galaxy and subsequently being accreted onto the Milky Way \citep[e.g.][]{malhan20}. 

\subsection{Generating an unperturbed stream} \label{codestep1}

The first step is to generate a model of an unperturbed AAU stream following the best-fit model in \citet{li21}. We use this unperturbed model of AAU since our goal is to create a mock of a stream with a similar morphology to the AAU stream. This stems from the idea that these stream subhalo encounters could make for a good candidate for a feature such as the AAU kink. 

We model our stream progenitor as a Plummer sphere \citep{Plummer_1911} with a mass of $\mathrm{2\times10^{4}\,M_{\odot}}$ and a scale radius of 10 pc. This progenitor matches fits to the AAU stream in \citet{li21}, which is a GC progenitor similar to those involved in the formation of the Palomar 5 stream \citep{erkal2017}. We generate the stream in the following gravitational potential: \\

$\bullet$ The Milky Way potential determined by \citet{mcmillan17} with accelerations defined using the C++ library of \texttt{galpot} \citep{dehnen98b}. This potential has 6 components; a bulge, spherical NFW dark matter halo \citep{navarro97}, thick and thin disk, and atomic and molecular hydrogen gas disks. Following the approach and results of \citet{li21,shipp21}, we use the same draw of the potential from the posterior chains of \citet{mcmillan17}. \citet{li21} explored several realizations and selected the one that gives the best fit to an unperturbed AAU stream. For the Sun's position, we use $\mathrm{R_0\,=\,8.22844\,kpc}$, and a velocity of $(U_\odot,\,V_\odot,\,W_\odot)\,=\,(8.40569,\,12.0149,\,7.28025)\,\mathrm{km\,s^{-1}}$ which are both specified by the draw from posterior chains in \citet{mcmillan17}. The parameters of this potential are given in Table A.3 of \cite{shipp21}.

$\bullet$ The Large Magellanic Cloud (LMC) is modeled as a Hernquist profile \citep{hernquist90} with a mass of $\mathrm{1.5\times10^{11}M_{\odot}}$ and a scale radius of $\mathrm{17.13\,kpc}$ as in \citet{li21} who fit the unperturbed portion of the AAU stream. Similar to that work, we choose this mass and scale radius which matches the LMC enclosed mass at $\mathrm{8.7\,kpc}$ \citep{marel14}. We note that this LMC mass is consistent with measurements from stellar streams, LMC satellites, and LMC globular clusters \citep[e.g.][]{erkal19,erkal2020,shipp21,vasiliev21,koposov23,watkins24}. We include the LMC alongside the Milky Way potential since it has a large effect on the AAU stream \citep[e.g.][]{shipp19,shipp21,li21} and so that the same model can be used to fit the real AAU stream in future work.

In order to generate the stream, we start with the present-day (i.e. $\mathrm{t\,=\,0}$) position and velocity of the progenitor, along with the LMC. The stream progenitor and the LMC are then integrated backward for a total of $\mathrm{4\,Gyr}$, which we will henceforth refer to as $\mathrm{t_{max}}$. While this integration takes place, the Milky Way position is not fixed due to the reflex motion on it exerted by the LMC \citep[e.g.][]{gomez15,erkal21}. This approach allows us to obtain positions for the LMC, MW, and our progenitor which can be used when integrating forwards from $\mathrm{t\,=\,-t_{max}}$. We use these positions when integrating forwards to avoid recomputing the forces again and interpolate their positions to suit our variable timestep.

The generation of the stream itself uses the modified Lagrange Cloud stripping method \citep{gibbons14} as adapted in \citet{erkal19} to include the LMC. We take a progenitor, modeled as a Plummer sphere and inject particles from the progenitor's inner and outer Lagrange points. This is done at randomly selected time intervals drawn from a Gaussian centered on each pericenter, with a mean of 2000 particles injected every pericenter. These particles are injected with velocities drawn from a Gaussian distribution chosen to match the velocity dispersion of the stream. The perturbed stars at the outer Lagrange point will therefore drift behind the progenitor and form the trailing arm of the stream, with those at the inner Lagrange drifting ahead and forming the leading arm. For our $\mathrm{4\,Gyr}$ integration to the present day, we end up with a stream of ~12,000 stars. We note that for the purpose of simulating the encounter, however, we do not integrate forwards to the present day in this initial step. Instead, we integrate until the time we intend to inject the subhalo for the encounter.

\subsection{Modelling the subhalo-stream encounter} \label{codestep2}
\begin{figure}
 \includegraphics[width=\columnwidth]{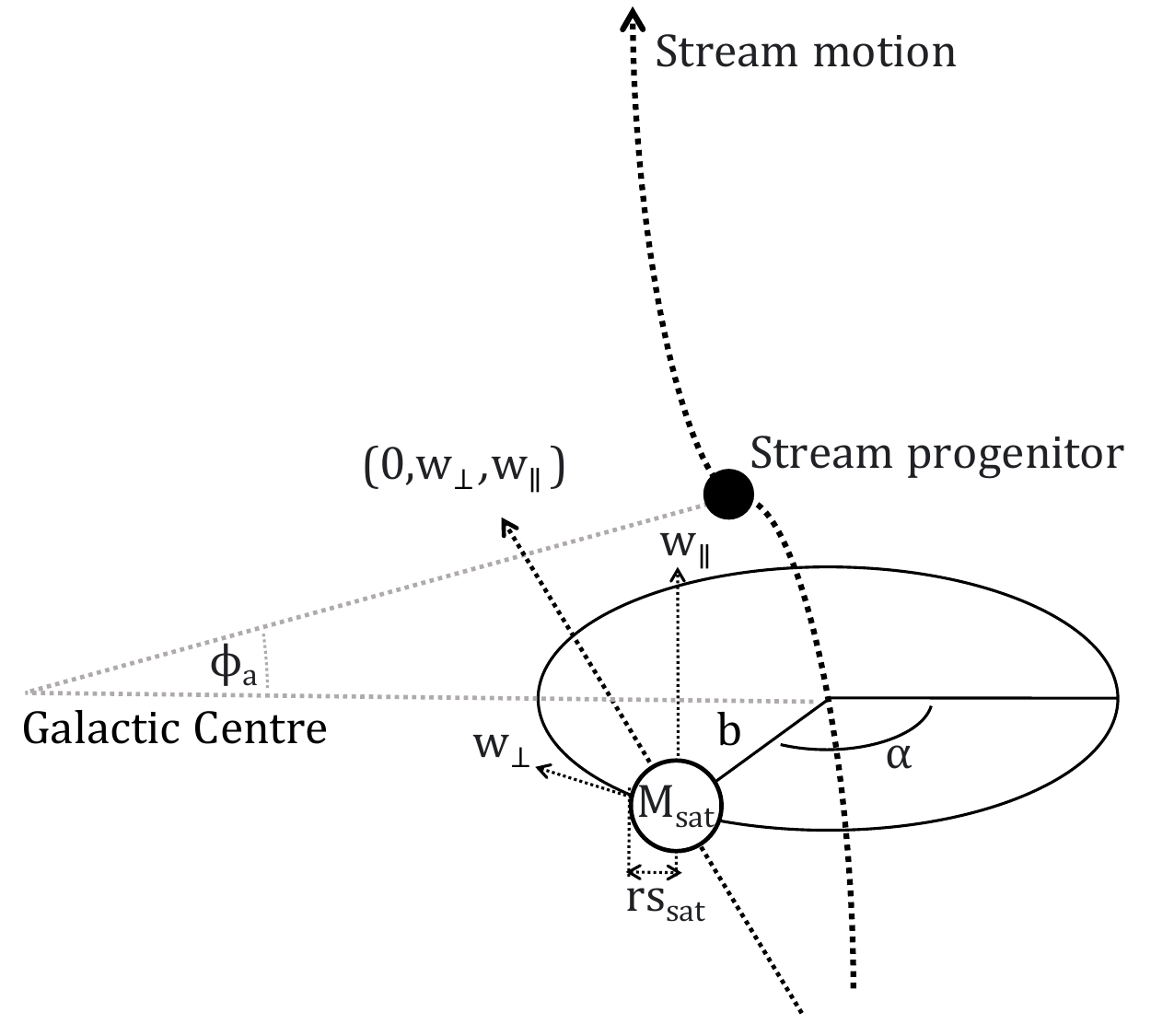}
 \caption{The setup of the Subhalo-Stream encounters, showcasing 7 of the 8 key impact parameters (the other being the time it took place before the present day, $\mathrm{T_a}$). The setup follows a cylindrical polar coordinate system, centered on the stream at angle $\mathrm{\phi_a}$}
 \label{fig:impactdiagram}
\end{figure}

Next, we describe how we initialize the subhalo properties given the unperturbed stream. Our encounters are modeled following the schematic shown in Figure \ref{fig:impactdiagram}. Using the stream produced in Section \ref{codestep1}, we use this to correctly align and position the subhalo with our impact parameters. In total, 8 parameters are used to describe these encounters, which are chosen to match the setup used in \citet{erkal15a}, and shown in Figure 1 in their respective paper.
The parameters are as follows: \\

$\mathrm{b}$: The impact parameter, i.e. the distance of the closest approach between the subhalo and stream. As a result, this is aligned so that it is roughly perpendicular to the stream and its motion.

$\mathrm{\alpha}$: An angle describing the orientation of the subhalo's closest approach to the stream. For a cylinder aligned along the stream, this angle would represent the cylindrical polar angle. This cylinder would run perpendicular to a line between the Galactic centre and the stream.

$\mathrm{w_\perp}$: The velocity of the subhalo perpendicular to the stream. This is aligned as a tangential velocity about $\mathrm{\alpha}$.

$\mathrm{w_\parallel}$: The velocity of the subhalo aligned parallel to the stream's velocity vector.

$\mathrm{T_{a}}$: The impact time, with the encounter set to occur at $\mathrm{T\,=\,-T_{a}}$

$\mathrm{\phi_{a}}$: The angle relative to the stream progenitor the encounter occurred. The other parameters relating to position and velocities are hence, aligned with this.

$\mathrm{M_{sat}}$: The subhalo mass

$\mathrm{rs_{sat}}$: The scale radius of the subhalo, which we assume to be a Plummer sphere \\

As the above parameters are defined relative to the position and velocities of the stream, we take a mean value of the stars local to where the encounter would be. At $\mathrm{\phi_{a}}$, we take a $0.5\,\deg$ slice to select only a local sample of stars for this. Once the subhalo can be placed with the relative positions and velocities, we then integrate it forward to the present day so that we can self-consistently rewind the progenitor in the presence of the subhalo in the next step. 

Similar to \citet{erkal15b}, we elect to fit the mass and scale radius of the subhalo independently (i.e. with no constraints on how they are related) as we do not want to restrict the possibility of these encounters to only certain dark matter paradigms. We note that for our fiducial impact, we select a subhalo mass and scale radius that match what is expected in $\Lambda$CDM.

\subsection{Generating the perturbed stream} \label{codestep3}

Finally, a stream is generated in the same process as Section \ref{codestep1}, though with the inclusion of the subhalo from the start of integration time. This is necessary so that we can self-consistently model the full effect of the subhalo, accounting for the orbit it is on and allowing for multiple impacts with the stream. We use the present-day positions of the subhalo and stream and rewind it to $\mathrm{t\,=\,-t_{max}}$, once again doing the same for the MW and LMC. This initial rewind means we can guarantee the stream ends up at the same present-day positions regardless of the subhalo properties, which may otherwise shift the orbit. This is one of the key advantages over carrying out the encounters using initial conditions at $\mathrm{t\,=\,-t_{max}}$. \\

\begin{figure*}
 \includegraphics[width=2\columnwidth]{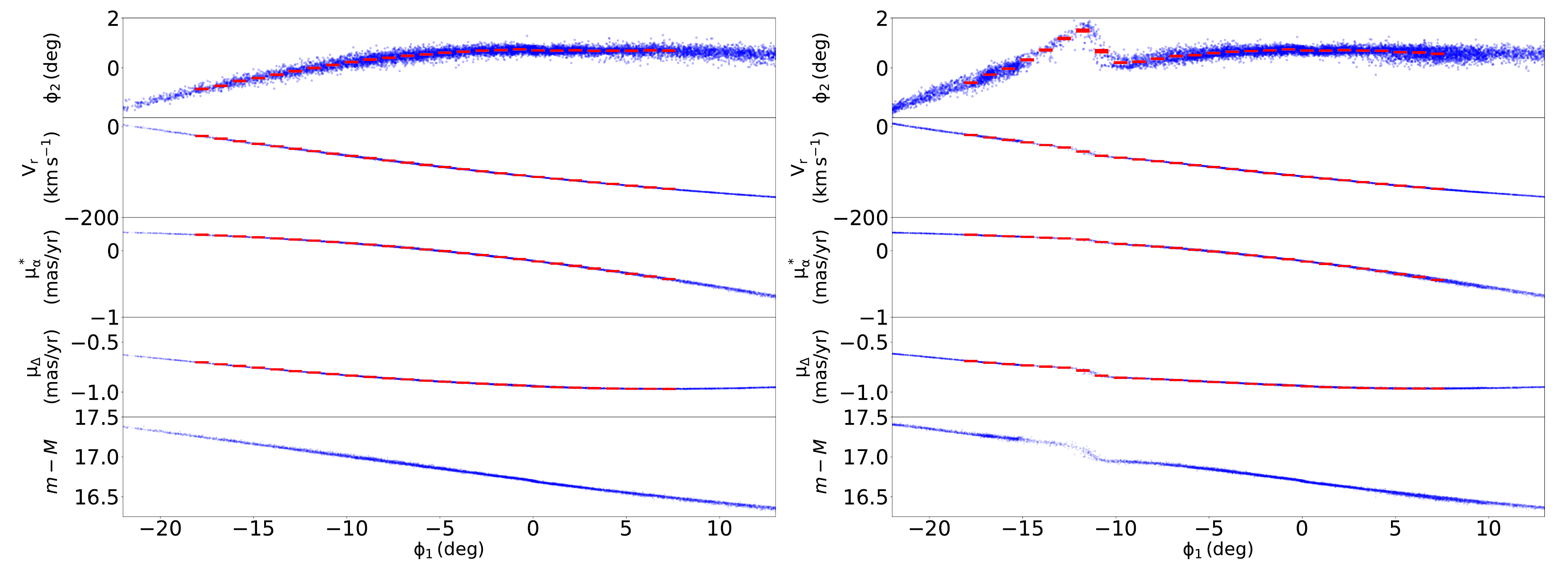}
 \caption{A comparison of a stream generated subject to only the background potentials and LMC as described in Section \ref{codestep1} (left) vs our mock stream that had additionally undergone a subhalo flyby (right). The exact parameters used here are $\mathrm{b\,=\,0.1\,kpc}$, $\mathrm{\alpha\,=\,250\,deg}$, $\mathrm{v_\perp\,=\,35\,km\,s^{-1}}$, $\mathrm{v_\parallel\,=\,-10\,km\,s^{-1}}$, $\mathrm{M_{sat}\,=\,10^7\,M_\odot}$, $\mathrm{T_a\,=\,0.25\,Gyr}$, $\mathrm{\phi_a\,=\,-6\,deg}$, $\mathrm{rs_{sat}\,=\,0.3\,kpc}$. The physical definitions of these parameters are outlined in figure \ref{fig:impactdiagram} and Section \ref{codestep2}. Both panels showcase the main stream observables in each row and these are outlined as follows: \textit{Top row:} Stream track, with errorbars every $\mathrm{1\,\deg}$ bin showing the mean $\mathrm{\phi_2}$ position of stars. \textit{Second row:} The radial velocities, with red errorbars in the same bins as the stream track showing mean radial velocities. Third and fourth row: The proper motion with similar errorbars for their respective bins. Bottom row: Distance modulus}
 \label{fig:unperturbedvsimpact}
\end{figure*}

\begin{figure}
 \includegraphics[width=\columnwidth]{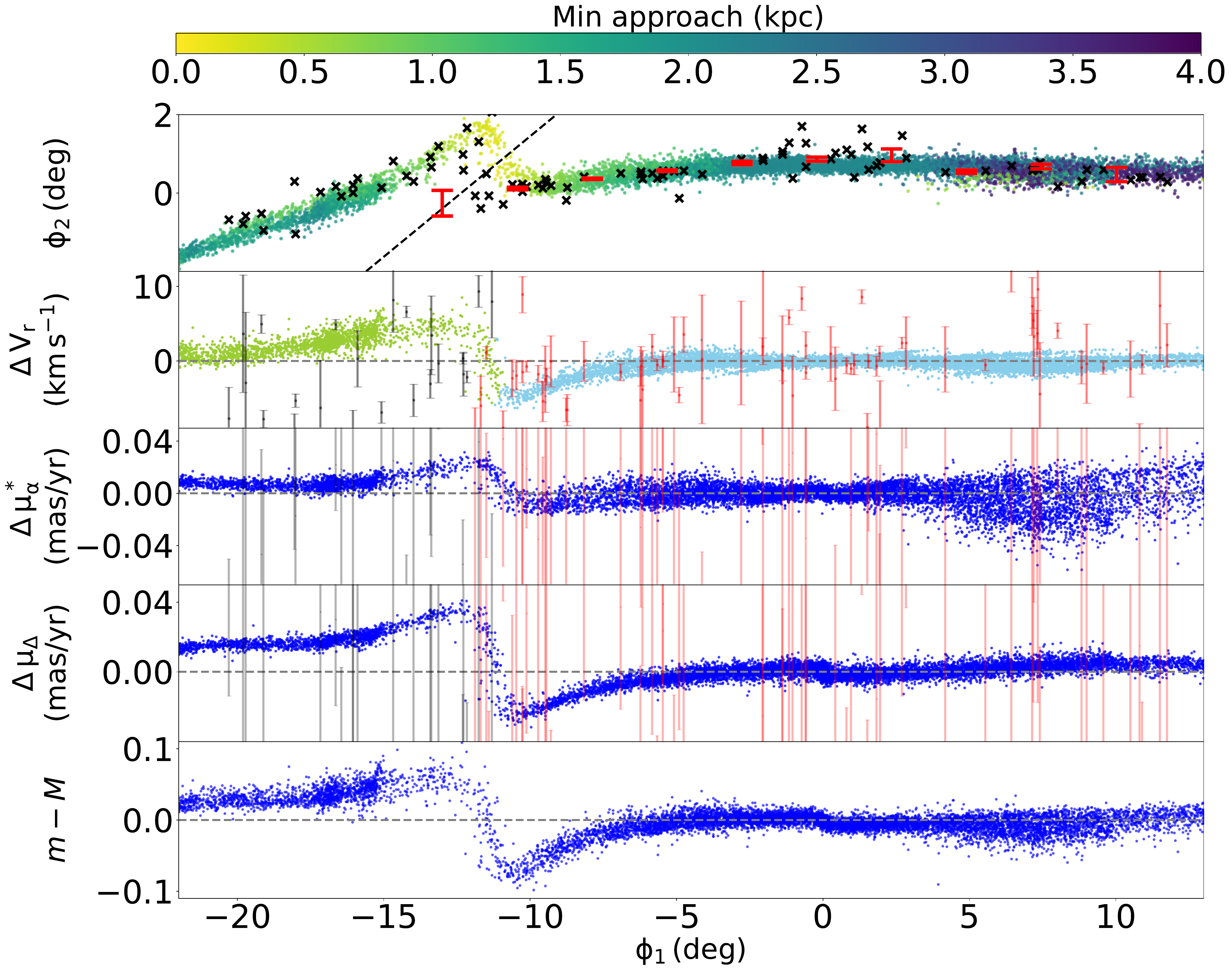}
 \caption{Mock data of our perturbed stream with real AAU data overlaid. \textit{Top row:} Stream track, with the minimum approach distance between individual mock stream stars and the subhalo shown in the colour bar. The black data points highlight the on-sky location of the observed stars. We also show a dotted line that separates the observed stars as belonging to either the ATLAS half of the stream (right of the kink) or the Aliqa Uma half (left of the kink). \textit{Second row:} The radial velocity differences between the mock perturbed stream and a fourth-order polynomial fit of $\mathrm{V_{GSR}}$ for the unperturbed model, with mock observables shown as red errorbars if in ATLAS and black errorbars if in Aliqa Uma. The model stars belonging to the ATLAS part of the stream are shown in light blue, and Aliqa Uma in green. Third and fourth row: The proper motion differences between the perturbed stream and a fourth-order polynomial fit, with mock observables shown once again as errorbars, much in the same vein as the radial velocities. Bottom row: Distance modulus difference between the perturbed stream and a fifth-order polynomial fit to the unperturbed model.}
 \label{fig:mockwithrealdata}
\end{figure}

\section{Fitting subhalo properties} \label{sec:recoverystuff}

Here, we showcase the encounter we use to generate the mock perturbed stream, along with the different sets of data we generate from this with the setups differentiated by the data quality. We then explain our approach to recovering subhalo properties from this mock data by determining the likelihood. 

We follow largely the same process as \citet{erkal15b} with a more realistic setup. In this work we use a realistic stream model instead of a train of particles on the same orbit which naturally accounts for the growth of the stream. Furthermore, we do not make use of an impulse approximation since our subhalo is integrated alongside the stream in the simulation. Finally, our setup allows us to model streams on eccentric orbits.

\subsection{The mock impact} \label{sec:mocksummary}

The exact parameters of our mock impact are as follows: $\mathrm{b\,=\,0.1\,kpc}$, $\mathrm{\alpha\,=\,250\,deg}$, $\mathrm{v_\perp\,=\,35\,km\,s^{-1}}$, $\mathrm{v_\parallel\,=\,-10\,km\,s^{-1}}$, $\mathrm{T_a\,=\,0.25\,Gyr}$, $\mathrm{\phi_a\,=\,-6\,deg}$, $\mathrm{M_{sat}\,=\,10^7\,M_\odot}$, $\mathrm{rs_{sat}\,=\,0.3\,kpc}$. These were chosen as the resulting stream track has a similar discontinuous morphology to the AAU stream with its kink feature. The same stream and observables, both with and without the subhalo encounter can be seen in Figure \ref{fig:unperturbedvsimpact}.

Additionally, we show how the current AAU data compares to this mock impact in Figure ~\ref{fig:mockwithrealdata}. This showcases features in not only track and radial velocities, similar to what was seen in \citet{li21}, but for proper motions. In the case of the proper motions, however, the scale of these seen in the mocks is far smaller than the errors of our current observations ($\lesssim 0.04$ mas/yr) so we would not expect to see this in AAU data. In addition, this mock has similar behaviour outside the kink where the rest of the track appears relatively unperturbed. 

We make mock observations in three different scenarios, corresponding to different uncertainties in the data. These setups have different observational errors and different numbers of stars in the observables of radial velocities and proper motions. In each scenario, we then generate our final set of mock data by drawing each mock observation from a Gaussian with their respective error distributions. 

We note, however, that in all cases we use the same errors for the stream track. In order to ensure that the random seed chosen to simulate the stream does not have a significant effect of the observables, we increase the mock uncertainty in the track by a factor of 5. The resulting observables have similar errors to present-day AAU observations along the stream track, and we, therefore, elect to use the same errors for data in these observables, only changing the quantity and location of data. Further details of each error scenario (hereafter, Scenario 1, Scenario 2, Scenario 3) are explained below.

\subsubsection{Scenario 1: No observational errors}

In this case, we use mock observations of the mean track of the stream in each observable and do not include any intrinsic uncertainties. Here, data is evenly spaced in $\mathrm{1\,\deg}$ bins in $\mathrm{\phi_1}$, and this is generated for each of the track, radial velocities, and both components of proper motions. In total, each observable has 26 data points between $\mathrm{\phi_1}$ of $\mathrm{-17.75^\circ}$ and $\mathrm{7.25^\circ}$.

\subsubsection{Scenario 2: Present day observational error}

In this setup, the quantity of data for observables and their errors closely match today's measurements of the AAU stream. The radial velocities and both proper motions individually consist of 96 stars, and are not evenly spaced throughout $\mathrm{\phi_1}$ like in the no error scenario. Instead, their positions in $\mathrm{\phi_1}$ match those of the AAU stream. We additionally modify the stream track data to be spaced more accordingly with our present-day spline fits in \citet{li21}. The result is 17 data points evenly spaced in $\mathrm{2\,\deg}$ bins in $\mathrm{\phi_1}$, with the total range being between between $\mathrm{-20^\circ}$ and $\mathrm{12^\circ}$.

\subsubsection{Scenario 3: Future predicted observational errors}

Looking forward, we investigate a mock up a future setup based on the expected performance of the Large Synoptic Survey Telescope (LSST), 4-metre Multi-Object Spectroscopic Telescope (4MOST), and the final data release of \textit{Gaia}. Here, our data for radial velocities now consist of 515 stars with an error ranging between $\sim$1 and $\sim$10$\mathrm{km\,s^{-1}}$, which would be enabled from 4MOST spectroscopic measurements to higher magnitude stars (r-mag of 21 compared to the current limit of 19.8) \citep{dejong12}. While LSST would be able to observe a far greater number of stars, the errors show little improvement to current AAU data (errors starting from $\sim2$ $\mathrm{mas\,yr^{-1}}$, which only improves upon approximately half of the current AAU proper motion errors) \citep{eyer12}. Our proper motions instead follow predictions from 10 years of \textit{Gaia} observations, where our dataset consists of 96 stars, but we improve on the present sample errors by a factor of $\sim$6.6 due to how the proper motions uncertainty scale as $\mathrm{t^{-1.5}}$ \citep{gaia23}. This gives a sample with errors ranging between $\sim$0.005 and $\sim$0.05 $\mathrm{mas\,yr^{-1}}$. We keep the stream track data the same as the present-day error scenario. 

\begin{figure}
 \includegraphics[width=\columnwidth]{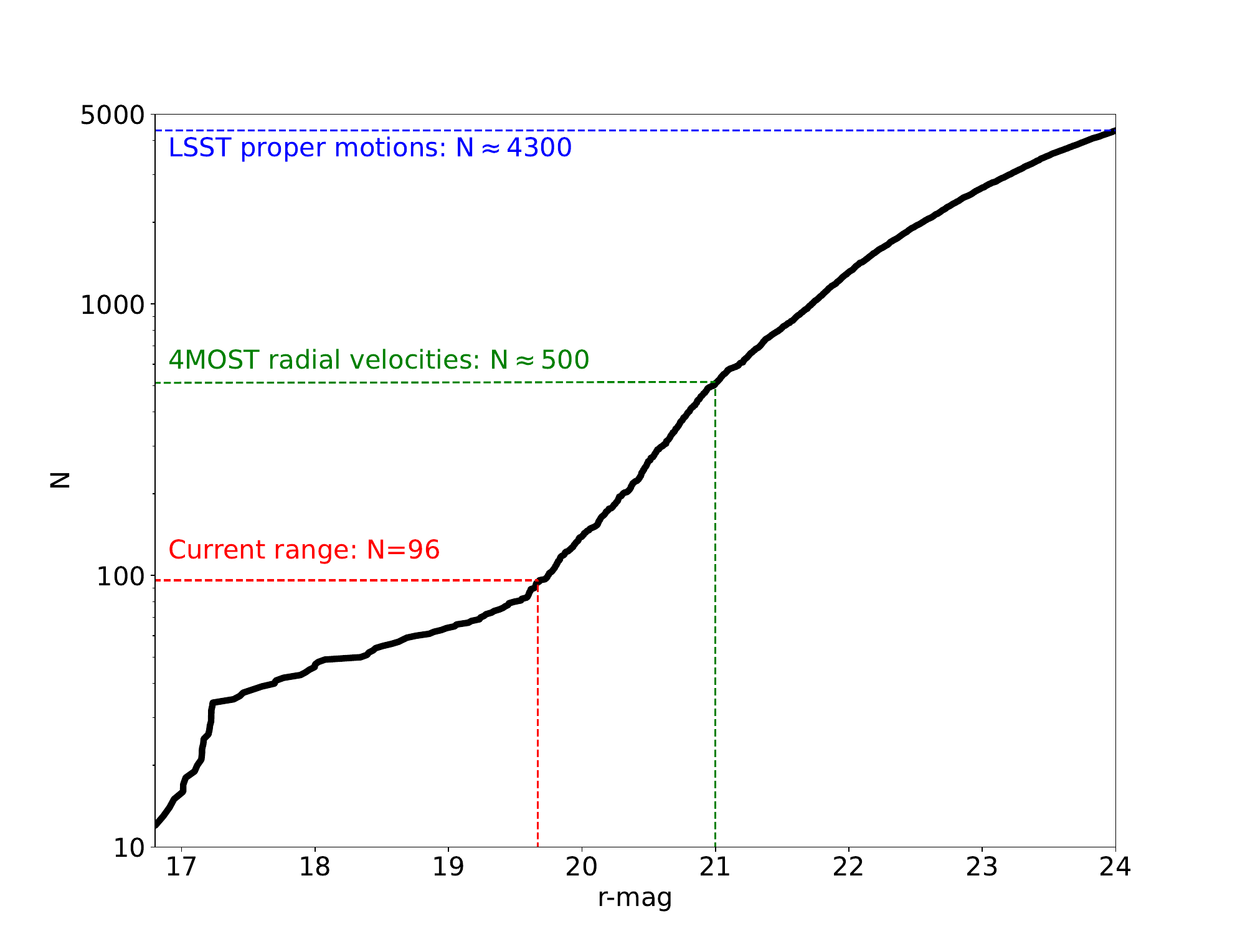}
 \caption{Estimates on the number of stars we would have data for as a function of magnitude. The current range described here shows the number of stars for which we have radial velocity and proper motion measurements in present-day AAU data, and we scale the IMF so that this luminosity function contains 96 stars in this range. The magnitude of stars that 4MOST and LSST will enable observations of are marked (21 mag for radial velocity measurements from 4MOST, and 24 mag for LSST proper motions), with the number of stars contained within the IMF in these ranges labeled. We note that we do not make use of LSST proper motions in our Scenario 3 fits and instead use proper motions expected from 10 years of \textit{Gaia} data.}
 \label{fig:numberiso}
\end{figure}

The mock data that would be expected from these future surveys was made by matching an isochrone to real AAU data, in a similar approach to \citet{erkal15b} and \citet{li21}. Here, this data was matched to the MIST isochrone sets \citep{dotter16, choi16} with the following parameters giving the best match; Stellar metallicity: $[\rm{Fe/H}\,=\,-1.99]$, $[\alpha/{\rm Fe}\,=\,0.4]$, Helium mass fraction: $[\mathrm{Y}\,=\,0.4]$ and age $\mathrm{\sim11.2\,Gyr}$. These were then interpolated using the \texttt{minimint} Python library \citep{koposovsoftware}. We use an initial mass function (IMF) following \citet{chabrier05}. This is scaled so that under the range covered by current AAU data ($14.8\leq \mathrm{r}\leq19.8$) the resulting luminosity function has the expected number of 96 stars contained within it, which equates to the current availability of radial velocity and proper motion data. The cumulative luminosity function of these stars is shown in Figure \ref{fig:numberiso}. This estimate results in a sample of 515 member stars that 4MOST will observe at $r<21$ mag.

\subsection{Likelihood Estimation}

We compute the log-likelihood of our encounters by assuming Gaussian errors for all observables. This gives a likelihood of

\begin{equation}
    \ln{\mathcal{L}}\,=\,\sum_{i}\left(\ln\left({\frac{1}{\sqrt{2\pi\sigma_{i}^{2}}}}\right)-\frac{(o_i-m_i)^2}{2\sigma_i^2}\right).
\end{equation}

\noindent where i denotes the index of the bin in $\phi_1$ along the stream, $o_i$ is the centroid for the simulated data in this bin, $m_i$ is the centroid for the model mock data in this bin, and $\sigma_i$ is the sum in quadrature of the uncertainty in the simulated data and the model mock data. This is carried out on the 4 observables, which are the stream track $\mathrm{\phi_2}$, radial velocities $\mathrm{\Delta\,v_{r}}$, and 2 proper motions, $\mathrm{\mu^{*}_{\alpha}}$ and $\mathrm{\mu_{\delta}}$. Similar to \citet{erkal15b}, we assume that these 4 observables are independent, so the total log-likelihood of a subhalo stream encounter is determined by simply summing the likelihood of the 4 individual observables. 

Given that Scenario 1 features no observational errors on individual measurements, the observational error here is represented as the uncertainty in the mean of a Gaussian fit to the simulated particles in each bin, i.e. the shot noise associated with having a finite number of particles. Likewise, the stream track for both Scenario 2 and Scenario 3 also has data points representing a bin of multiple stars (17 data points in the track between $\mathrm{-20^\circ}$ and $\mathrm{12^\circ}$). Therefore, for every observable in Scenario 1, alongside the stream track for Scenario 2 and Scenario 3, the error in our simulated data is simply represented by the error in the mean for each bin. The radial velocities and proper motion data for both Scenario 2 and Scenario 3 meanwhile represent individual observations of stars. Here, the error in simulated data we use is instead the spread in the data of simulated stars about each bin.

\begin{figure}
 \includegraphics[width=\columnwidth]{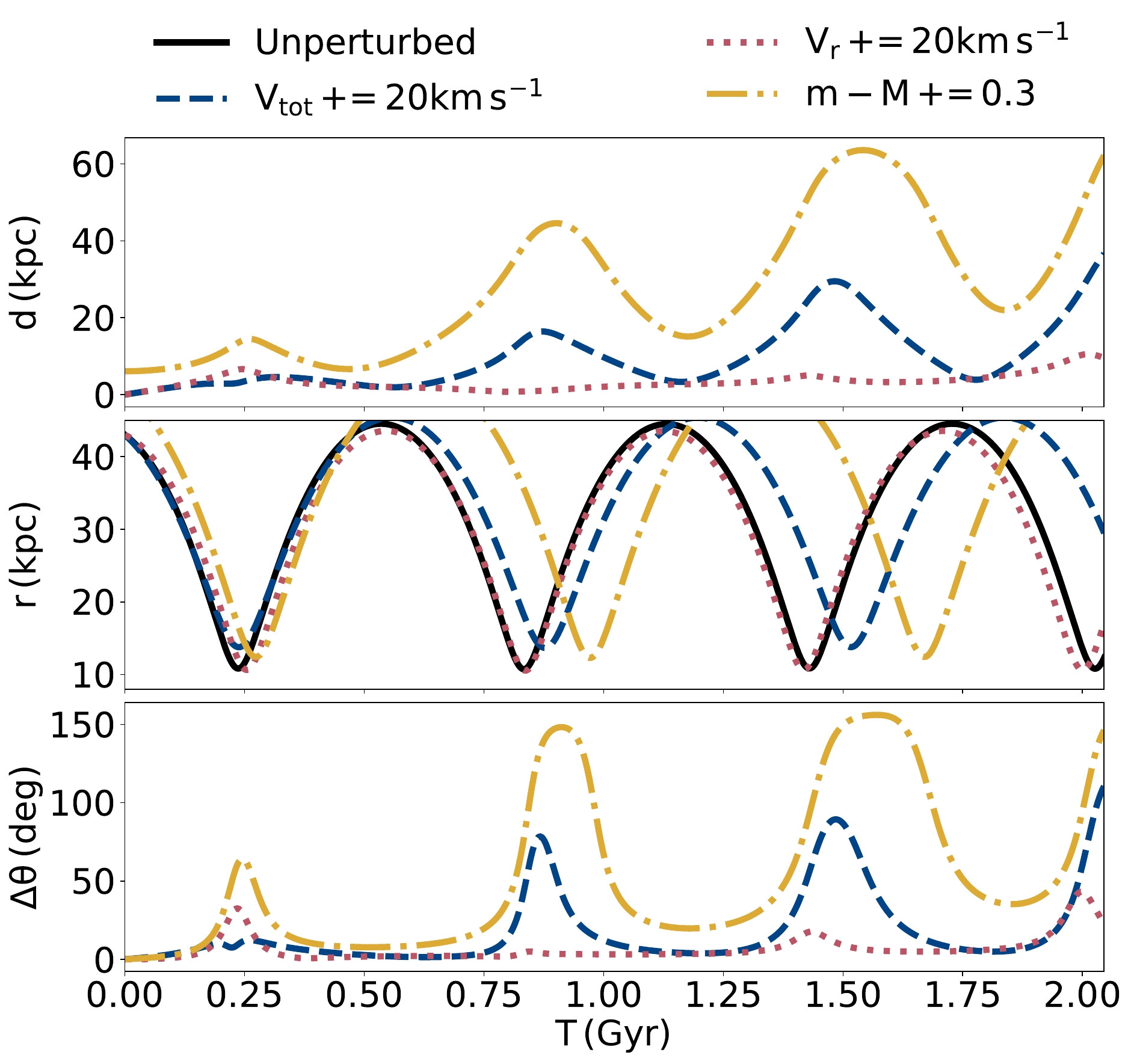}
 \caption{The evolution of particle positions for a total of $\mathrm{2\,Gyr}$ after applying a perturbation to a duplicate of a particle during stream generation. 3 different perturbations were applied individually here; Red: An increase in velocity along the stream of $\mathrm{20\,km\,s^{-1}}$, Blue: An increase in the radial velocity in galactic coordinates of $\mathrm{20\,km\,s^{-1}}$, Green: An increase in the distance modulus of 0.3. The panels are arranged as follows; Top: The distance between the perturbed and unperturbed particle, Middle: The galactocentric radius of both the perturbed and unperturbed particles, Bottom: The difference in sky positions between the perturbed and unperturbed particles.}
 \label{fig:distances}
\end{figure}

\subsection{A recent impact time}

Our impact time for the mock uses a subhalo which encountered the stream fairly recently (250$\,\mathrm{Myr}$). The goal is to generate a stream with morphology similar to the AAU kink. We expect this flyby to apply velocity kicks to stars very close to each other along the track in $\mathrm{\phi_1}$. A small velocity perturbation, or an offset in distance modulus, would result in different orbital periods for these stars which would cause their position along the stream to change drastically over time as they drift from these original positions. As a result, we find that older encounters are not able to create a similar-looking mock stream to the AAU kink. 

We test this using a simple approach of selecting a single stream particle approximately 2 Gyr after stream generation began. We duplicate this particle and apply a kick in one of this duplicate's velocity along the stream, $\mathrm{V_{tot}}$, the radial velocity relative to us, $V_{r}$, or a change in its distance modulus, $m\,-\,M$. The size of these are $20\,{\rm km\,s}^{-1}$ in velocities, and 0.3 in distance modulus. We then follow these two particles forward in the stream for 2 Gyr, shown in Figure \ref{fig:distances}. We can clearly see that shortly after roughly 0.6 Gyr, the particles are each separated by some tens of degrees, which would leave little trace of any feature especially any similar to a kink. While this is a qualitative look at the nature of the encounter, for the purpose of this work we assume this is the most plausible nature surrounding a subhalo-stream encounter that could produce streams with discontinuous morphology such as AAU. The high dimensionality of the space however could mean that further exploration has the potential to reveal different, older impacts that cause similar features.

\section{MCMC Analysis} \label{sec:MCMCresults}

In order to explore how well the subhalo properties can be constrained, we use a Markov Chain Monte Carlo (MCMC) sampler to explore our likelihood space. The MCMC is a useful tool for intuitively exploring parameter space. Here, we use the Python library \textsc{emcee} \citep{2013PASP..125..306F} for our suite of MCMCs. For all of our runs, we use the same priors and they are summarised in Table \ref{tab:priors}. \\

\begin{table}
	\centering
	\caption{An overview of the parameters and priors for our MCMC setups. The descriptions of each parameter are covered in Section \ref{codestep2}, however, here we represent the priors on $\mathrm{w_\perp}$ and $\mathrm{w_\parallel}$ by the combined, total velocity of the subhalo $\mathrm{V_{tot}}$.}
	\label{tab:priors}
	\begin{tabular}{lccr}
		\hline
		Parameter & Distribution & Prior\\
		\hline
		b & Uniform & $\mathrm{(0,\inf)\,kpc}$\\
		$\mathrm{\alpha}$ & Uniform & $\mathrm{(0,360)\,\deg}$\\
		$\mathrm{V_{tot}}$ & Maxwellian & $\mathrm{(168.2\,\pm\,97.1)\,km\,s^{-1}}$\\
		$\mathrm{T_a}$ & Uniform & $\mathrm{(0,4)\,Gyr}$\\
		$\mathrm{\phi_a}$ & Uniform & $\mathrm{(-180,180)\,\deg}$\\
		$\mathrm{M_{sat}}$ & Uniform & $\mathrm{(0,\inf)\,M_\odot}$\\
		$\mathrm{rs_{sat}}$ & Uniform & $\mathrm{(0,\inf)\,kpc}$\\
		\hline
	\end{tabular}
\end{table}

These priors are based on previous works that modeled subhalo impacts \citep[][]{yoon11,Carlberg_2012,erkal16}. These priors are based on the expected distribution of the closest approaches of subhaloes, which are assumed to be uniformly distributed in position, with a line, which represents the stream \citep[e.g. see Sec. 2 of][]{erkal16}. 

We note that our prior of a Maxwellian distribution on our total velocity also serves to address the mass-velocity degeneracy that was encountered in \cite{erkal15b}. In their analytic work, they assumed the impulse approximation which means that the stream particle and subhalo are assumed to move on straight lines during the encounter. Furthermore, their model assumes that the velocity kicks are imparted instantaneously at the time of closest approach. In their model, the velocity kicks imparted by the subhalo are proportional to the subhalo mass divided by the flyby speed. Thus the mass and subhalo speed can be arbitrarily scaled and yield exactly the same result. A priori, it is unclear how strong this degeneracy should be in our more realistic setup since we include the subhalo as a particle orbiting in the host potential with the stream. This accounts for the actual orbit of a stream and the subhalo and also accounts for the fact that the acceleration of the subhalo on the stream is not instantaneous. This is especially important at lower relative velocities where the accelerations on stars would be experienced over a very significant period of time instead of instantaneously. Still, similar to their work we include a Maxwellian distribution applied to the total subhalo velocity. Here, we use a dispersion in the total subhalo velocity, $\mathrm{\sigma}$, which relates to the circular velocity at $\mathrm{10\,kpc}$, $v_{\mathrm{circ}}$, following the expression $\mathrm{\sigma\,=\,v_{circ}/\sqrt{3}}$. This results in $\mathrm{\sigma\,=\,97.1,km\,s^{-1}}$. 

Next, we present the MCMC results for all 3 data quality scenarios. For every scenario, our results shown involve an MCMC using 100 walkers, and we burn-in 1/2 of all steps.

\subsection{Scenario 1: Fits with no observational errors} \label{sec:result_no_error}

\begin{figure*}
 \includegraphics[width=2\columnwidth]{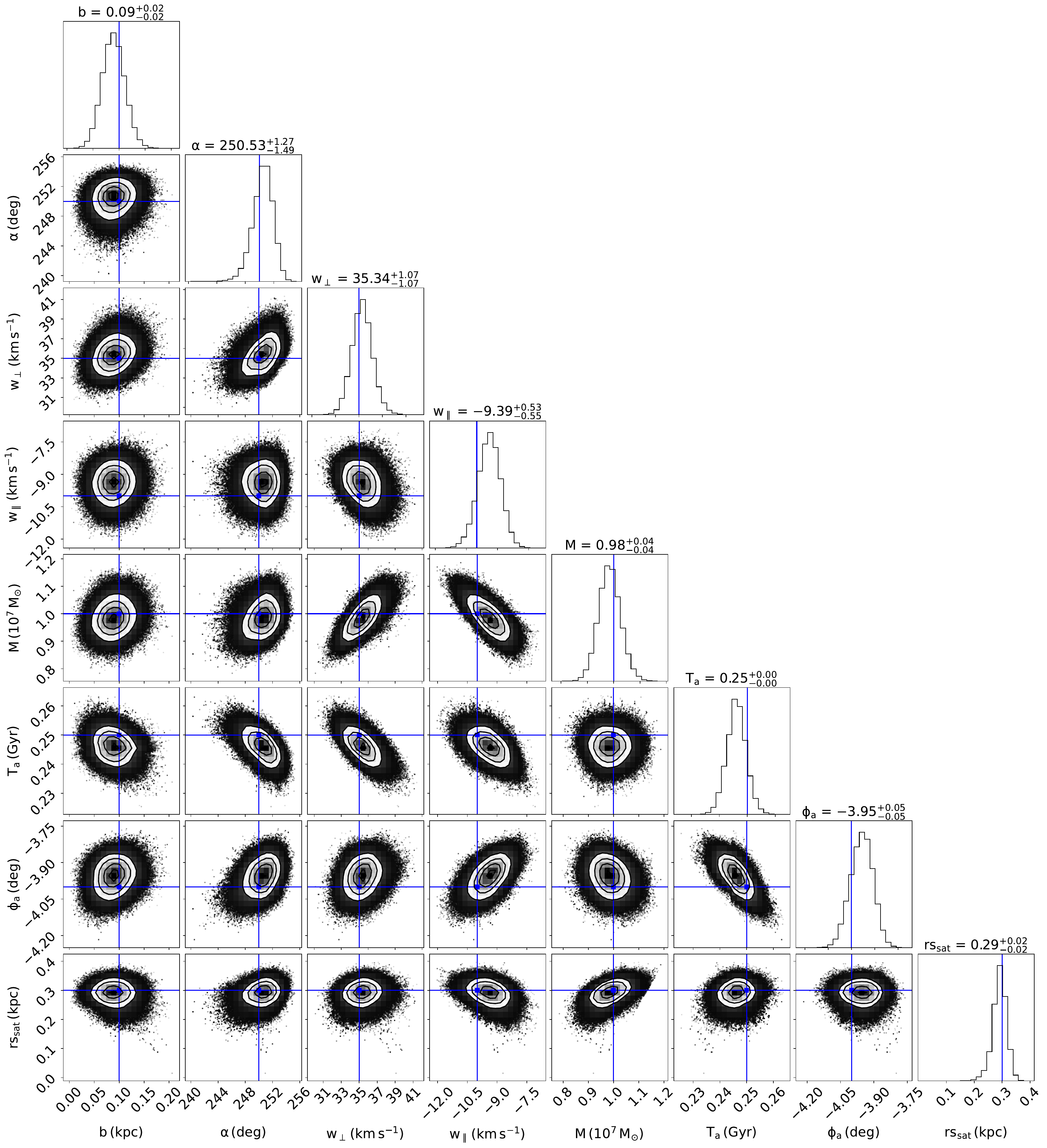}
 \caption{Corner plot of $\sim$40,000 steps of our MCMC in the scenario using no observational errors. The results here are shown after burning half of the steps. All 8 impact parameters are shown, as summarised in Section \ref{codestep2}, with their corresponding 1d histograms along the top of the plot. The contour lines here represent the 3 sigma levels of the steps}
 \label{fig:noerrorcorner}
\end{figure*}

First, we consider our approach with no additional observational errors. We evolved our MCMC for 38,370 MCMC steps and show the results in Figure \ref{fig:noerrorcorner}. The autocorrelation times are $\sim$600 for each variable, so this number of steps was sufficient to reach more than 50 autocorrelation times. For every panel, the true parameters were recovered within at least 2 sigma, which shows that the parameters could be inferred well for Scenario 1. The likelihood peaks in the figure do not coincide exactly with the true values. This is expected since our mock measurements have been sampled using their respective errors as explained in Section \ref{sec:mocksummary}. 

We note that some of the correlations in Figure \ref{fig:noerrorcorner} are elongated showing some slight degeneracies exist. When comparing either velocities, $\mathrm{w}$, with $\mathrm{M_{sat}}$, this is fairly evident, where the high likelihood solutions are stretched along $\mathrm{M_{sat}} \propto \mathrm{w}$. This degeneracy was present in \citet{erkal15a} as an unbroken degeneracy, i.e. arbitrary rescalings of $\mathrm{w}$ and $\mathrm{M_{sat}}$ led to exactly the same gap. We note that in our setup this degeneracy is broken by the fact that we do not make an impulse approximation and instead integrate the subhalo in the host potential along with the stream. In addition, we do have a prior on the velocity although this prior is much weaker than the constraints shown in Figure \ref{fig:noerrorcorner}. There is also a similar feature between $\mathrm{T_a}$ and $\mathrm{\phi_a}$. This weak degeneracy arises due to the fact the stream is on an eccentric orbit and thus stretches and compresses as it orbits the Milky Way. As a result, if the impact time is slightly later or earlier, the subhalo impact point along the stream will have to be moved slightly to have the same present-day location of the kink. 

\subsection{Scenario 2: Fits using current observational uncertainties}

Next, we show the results of Scenario 2 with its present-day observational uncertainties. The resulting chains are shown in Figure \ref{fig:presenterrorcorner}. In this case, we run our MCMC for 61,142 steps. We note that despite this large number of steps, since the largest autocorrelation times are roughly 2,000, this does not reach the target of 50 times the autocorrelation times. In general, the parameters were still well recovered, with most being within 2 sigma of their true values once again. Similar to Scenario 1, the velocity with mass panels alongside the impact time with angle along the stream exhibit faint hints of degeneracies present. However, in the latter case, this is still not much more significant than Scenario 1 in Figure \ref{fig:noerrorcorner}. 

The recovery of $\mathrm{w_{\perp}}$ and $\mathrm{M_{sat}}$ was not as successful, where the true value for the mock was over 3 sigma off the results. Since these parameters are strongly correlated and the true values of $\mathrm{w_{\perp}}$ and $\mathrm{M_{sat}}$ lie along the correlation, this is likely a result of both the degeneracies between mass and velocity combined with our priors and the larger volume in parameter space at higher velocities. The true impact parameters do give a lower velocity than the predicted circular velocity from our Maxwelian distribution, and this would result in their likelihood being disfavoured slightly compared to high-velocity encounters. It is possible even excluding this prior, these higher velocities and masses still give very similar likelihoods compared to the true impact.

\subsection{Scenario 3: Fits using future observational uncertainties}

Finally, we consider our future error scenario. Here, our chains are run for a total of 85,140 steps with the posteriors shown in Figure \ref{fig:futureerrorcorner}. As with Scenario 2, this is lower than the target of having steps equal to 50 times the autocorrelation times, which in this instance were on average roughly 5,000 as well. The parameters are still well recovered, with most being within 2 sigma of their true values. Compared to Scenario 2, the recovered parameters are closer to the true values and the uncertainties are significantly smaller. We also note the same degeneracies that were seen in both previous error scenarios, albeit with a significantly smaller uncertainty compared to Scenario 2. 

\begin{figure*}
 \includegraphics[width=2\columnwidth]{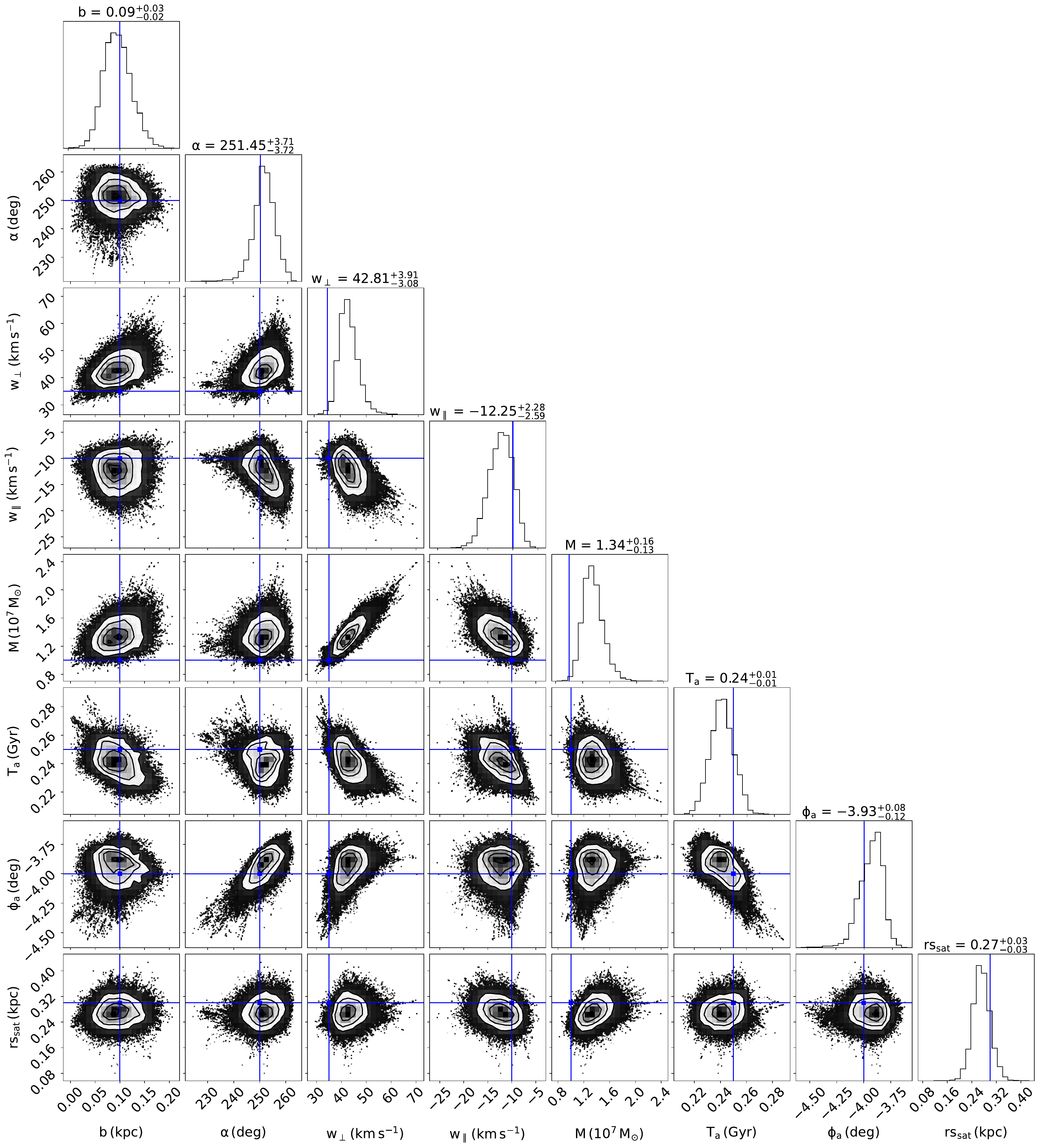}
 \caption{Corner plot of ~60000 steps of our MCMC in the scenario using present-day observational errors. The results here are shown after burning half of the steps. All 8 impact parameters are shown, as summarised in Section \ref{codestep2}, with their corresponding 1d histograms along the top of the plot. The contour lines here represent the 3 sigma levels of the steps}
 \label{fig:presenterrorcorner}
\end{figure*}

\begin{figure*}
 \includegraphics[width=2\columnwidth]{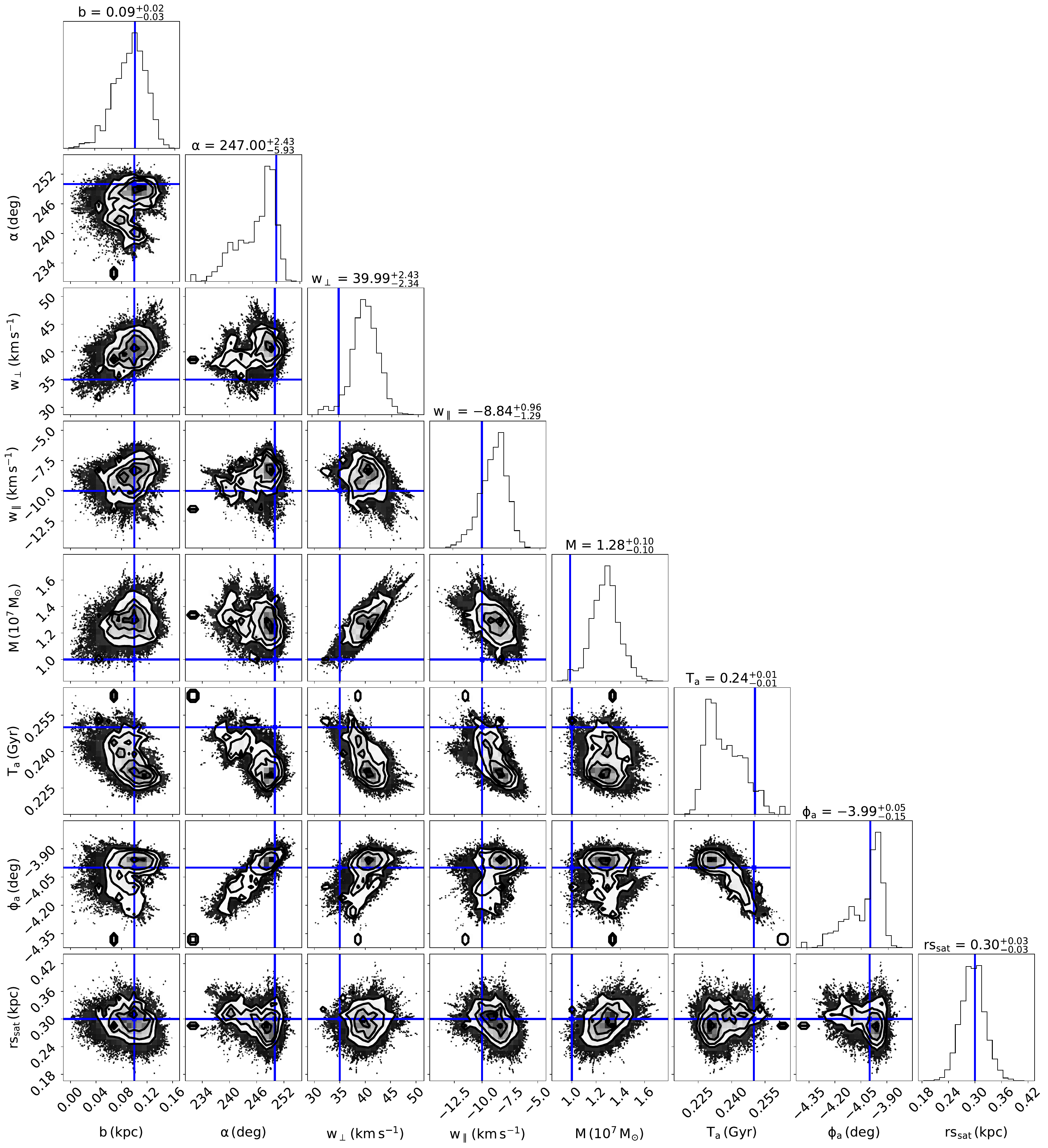}
 \caption{Corner plot of ~80000 steps of our MCMC in the scenario using future observational errors. The results here are shown after a burn-in of 1/2 of all steps. All 8 impact parameters are shown, as summarised in Section \ref{codestep2}, with their corresponding 1d histograms along the top of the plot. The contour lines here represent the 3 sigma levels of the steps}
 \label{fig:futureerrorcorner}
\end{figure*}

We note that both the subhalo mass and the speed of the subhalo are underestimated. As in Scenario 2, this is likely due to a partial degeneracy between the mass and speed of a subhalo, combined with our prior and the larger volume in our parameter space at higher speeds. In addition, while the impact time $\mathrm{T_{a}}$ is still recovered, there is an interesting asymmetry in the posterior. The correlation between $\mathrm{T_a}$ and $\mathrm{\phi_a}$ is  more significant compared to Scenario 1 and 2 which appears to be driving this distribution. We also note that while the inferred parameters in Scenario 3 appear to have a lot of noise compared to Scenario 2 (e.g. compare Fig. \ref{fig:futureerrorcorner} and \ref{fig:presenterrorcorner}), the ranges of all of the parameters are significantly smaller so the parameters are recovered with a much higher precision. 

\begin{figure}
 \includegraphics[width=\columnwidth]{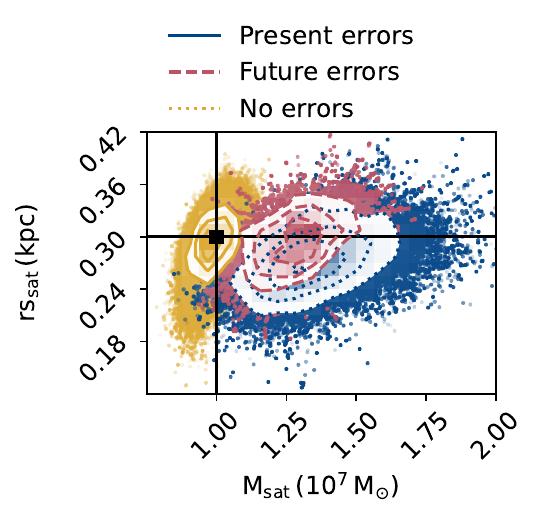}
 \caption{The recovery of the mass and scale radius of the subhalo for each of the 3 error scenarios. Yellow: No observational errors, Blue: Present-day observational errors, Red: Future predicted observational errors. The contours for each scenario showcase the 3 sigma levels, and the scatter shows the full range of parameters explored after burning half of all steps. The true mass and scale radius values for the mock impact are shown in black}
 \label{fig:mass_scaleradius}
\end{figure}

\begin{figure}
 \includegraphics[width=\columnwidth]{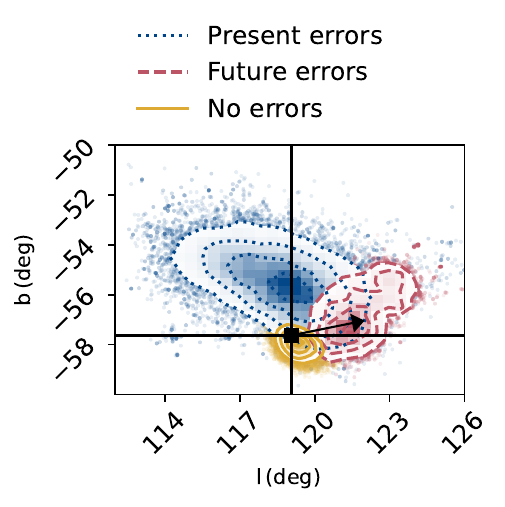}
 \caption{The recovery of the sky positions of the subhalo in the present day for 2500 randomly sampled MCMC runs for all 3 scenarios. Yellow: No observational errors, Blue: Present-day observational errors, Red: Future predicted observational errors. The contour levels represent the sigma levels for all 2500 steps, and the scatter shows the full range of possible subhalo positions from the subhalo-stream encounters used. The subhalo position of the mock impact is shown in black, with its direction of orbit also shown.}
 \label{fig:subhalopositions}
\end{figure}

\section{Discussion} \label{sec:discussion}

\subsection{Subhalo Localisation}

Given that the fits can reliably determine the subhalo properties at impact, we explore how well they can localize the subhalo at the present day. For each error scenario, we randomly chose 25000 encounters from the MCMC samples after burn-in of 1/2 to follow their orbits during and after the encounter. Specifically, we obtained the subhalo positions at $\mathrm{t=0}$, mimicking their expected locations on the present day. In Figure \ref{fig:subhalopositions} we show how well the sky positions of the subhalo can be constrained for each scenario, additionally showing the position and motion of the mock subhalo. This is also shown in galactocentric coordinates in Figure \ref{fig:subhalopositionsXYZ}. Even for the largest errors of Scenario 2, we can expect a relatively tight constraint on its location, which shows great potential for future fits on real observational data. This relatively small area covered will be in part due to the recent impact time, where differences in subhalo velocity between fits will only translate to a minor change in present-day sky positions. Future fits on streams, particularly if the impact time was many orbital periods ago, could show a far broader range in present-day subhalo positions.

If the subhalo's present-day position is well-localized, this can help with searching for possible baryonic perturbers. For example, globular clusters and dwarf galaxies can also perturb stellar streams and the subhalo's present-day position could be compared with known objects. Similarly, the subhalo's present-day location could be compared with maps of gamma-ray emission to each for possible evidence of dark matter annihilation or decay \citep[e.g.][]{Mirabel_Bonaca_2021}.

\subsection{Information content of perturbed streams}

In this work, we have shown that the properties of the impacting subhalo (i.e. its mass, scale radius, impact position and velocity, and impact time) can be recovered from observations of a perturbed stellar stream. Previously, \cite{erkal15b} had shown similar results for streams on circular orbits. In that work, they used an analytic model for a perturbed stream which assumed that the stream was a train of particles on the same (circular) orbit. In addition, they also assumed that the subhalo impact was impulsive, i.e. that the stream and subhalo moved on straight lines during the encounter. In that work, they explored the information content of the observations and showed that all parameters describing the subhalo's encounter with the stream could be recovered up to a degeneracy between the subhalo mass and its velocity relative to the stream. As shown in Section \ref{sec:result_no_error}, this degeneracy is not as prevalent in our work since we do not assume the impulse approximation and instead inject the subhalo on its orbit. We note that \cite{bonaca19} also found a similar result in their fits to the GD-1 stream.

Another relevant result of \cite{erkal15b} is that they proved that the subhalo parameters could be extracted using measurements of only three stream observables, e.g. stream track on the sky, radial velocity, and one proper motion. In this work, we used four observables (stream track, radial velocity, and two proper motions) and were able to get robust measurements of all the subhalo properties. This is similar to the setup discussed in Section 5.3.3 of \cite{erkal15b} which emphasized that the subhalo properties can be inferred without any density information. This is a useful approach since it does not require that we understand the disruption history of the globular cluster progenitor which can affect the density profile \citep[e.g.][]{Gieles+2021}. Thus, the results of this work show that the intuition from \cite{erkal15b} should extend to realistic streams on eccentric orbits.

\section{Conclusions} \label{sec:conclusions}

In this work, we have explored how well the properties of subhaloes can be recovered from observations of stellar streams in a realistic setup. Using synthetic data sets with different amounts of errors, we modeled subhalo-stream encounters in a Bayesian framework and inferred how well we can constrain the properties of these encounters in numerical simulations. We first modeled these setups and described the encounters using 8 key parameters. Then, we focused on generating perturbed streams with similar morphology to the kink in the AAU stream and did so by running a perturbation featuring an almost direct impact of a $\mathrm{10^7\,M_\odot}$ subhalo onto the stream. We summarise our conclusions below: \\

$\bullet$ While a stream morphology similar to the AAU kink can be reproduced by subhalo-stream encounters, including producing $\sim 10$ km/s velocity differences across the kink, these encounters appear to require very recent impacts. This is due to the rate at which particles drift apart after receiving a kick in velocity or distance modulus, which would very rapidly drift apart tens of degrees after $\sim0.6$ Gyr and leave little trace of any feature.

$\bullet$ In instances where we have effectively no observational errors, we can accurately recover the full suite of 8 impact parameters defining the encounter. This gives us full dynamical data of the subhalo relative to the stream, as these include not only the velocities but also where and when the encounter took place. In addition, we can recover the subhalo mass and scale radius.

$\bullet$ If we include realistic observational uncertainties, there usually are difficulties recovering the velocities of the subhalo and its mass. This is in part due to the mass velocity degeneracy still partially present even in numerical simulations. Still, these tend to be recovered within 3 standard deviations for the worst cases and shows promise in breaking this often encountered degeneracy in previous works \citep[e.g.][]{erkal15b}.

$\bullet$ The recovery above is only minimally reduced by improving the observational errors and adding to the abundance of data as well, and in the case of some other parameters, there is no improvement at all to the recovery. Still, for these other parameters, both maintain a recovery within one standard deviation just like the no-error scenario. 

$\bullet$ Of the properties, the mass and scale radius carry a lot of future potential if this could be applied to fitting real streams. These two parameters can be used to test how well the properties of low-mass dark matter subhaloes compare with predictions from cosmological simulations with cold dark matter and alternative dark matter models.

$\bullet$ Regardless of the quality of the observables, we find that we can constrain the present-day sky positions of the subhalo within ~$\mathrm{10\deg}$ for this recent impact. We can use this for real impacts to perform observations in this region for any luminous objects that may fit the properties of our subhalo, to search for possibly baryonic perturbers, and to search for possible gamma-ray signatures of dark matter annihilation.

The results of this work suggest that it should be possible to use perturbations like those present in AAU to measure the perturber which created them. Furthermore, such fits should already be possible with current observational uncertainties and these fits should improve in precision with upcoming data from surveys like 4MOST and further \textit{Gaia} data releases. As more perturbed streams are observed, fits like these will allow us to build a census of perturbers in the Milky Way. While some of these may be due to baryonic substructure, e.g. due to a dwarf galaxy, some will hopefully have no baryonic culprit and will instead allow us to build an inventory of dark matter subhaloes in the Milky Way.

\section*{Acknowledgements}

DE acknowledges support through ARC DP210100855.  SK acknowledges support from the Science \& Technology Facilities Council (STFC) grant ST/Y001001/1.
T.S.L. acknowledges financial support from Natural Sciences and Engineering Research Council of Canada (NSERC) through grant RGPIN-2022-04794. A.P.J. acknowledges support by the National Science Foundation under grants AST-2206264 and AST-2307599. DZ and GFL acknowledge support from the Australian Research Council Discovery Project DP220102254. N.S. is supported by an NSF Astronomy and Astrophysics Postdoctoral Fellowship under award AST-2303841. G.L. acknowledges FAPESP (Proc. 2021/10429-0).

For the purpose of open access, the author has applied a Creative
Commons Attribution (CC BY) licence to any Author Accepted Manuscript version
arising from this submission.

\section*{Data Availability}

The models and chains in this work will be made available upon reasonable request to the corresponding author.



\bibliographystyle{mnras}
\bibliography{example} 

\begin{thebibliography}{}
\makeatletter
\relax
\def\mn@urlcharsother{\let\do\@makeother \do\$\do\&\do\#\do\^\do\_\do\%\do\~}
\def\mn@doi{\begingroup\mn@urlcharsother \@ifnextchar [ {\mn@doi@} {\mn@doi@[]}}
\def\mn@doi@[#1]#2{\def\@tempa{#1}\ifx\@tempa\@empty \href {http://dx.doi.org/#2} {doi:#2}\else \href {http://dx.doi.org/#2} {#1}\fi \endgroup}
\def\mn@eprint#1#2{\mn@eprint@#1:#2::\@nil}
\def\mn@eprint@arXiv#1{\href {http://arxiv.org/abs/#1} {{\tt arXiv:#1}}}
\def\mn@eprint@dblp#1{\href {http://dblp.uni-trier.de/rec/bibtex/#1.xml} {dblp:#1}}
\def\mn@eprint@#1:#2:#3:#4\@nil{\def\@tempa {#1}\def\@tempb {#2}\def\@tempc {#3}\ifx \@tempc \@empty \let \@tempc \@tempb \let \@tempb \@tempa \fi \ifx \@tempb \@empty \def\@tempb {arXiv}\fi \@ifundefined {mn@eprint@\@tempb}{\@tempb:\@tempc}{\expandafter \expandafter \csname mn@eprint@\@tempb\endcsname \expandafter{\@tempc}}}

\bibitem[\protect\citeauthoryear{{Banik}, {Bovy}, {Bertone}, {Erkal}  \& {de Boer}}{{Banik} et~al.}{2021a}]{Banik_etal_2021a}
{Banik} N.,  {Bovy} J.,  {Bertone} G.,  {Erkal} D.,   {de Boer} T.~J.~L.,  2021a, \mn@doi [\mnras] {10.1093/mnras/stab210}, \href {https://ui.adsabs.harvard.edu/abs/2021MNRAS.502.2364B} {502, 2364}

\bibitem[\protect\citeauthoryear{{Banik}, {Bovy}, {Bertone}, {Erkal}  \& {de Boer}}{{Banik} et~al.}{2021b}]{Banik_etal_2021b}
{Banik} N.,  {Bovy} J.,  {Bertone} G.,  {Erkal} D.,   {de Boer} T.~J.~L.,  2021b, \mn@doi [\jcap] {10.1088/1475-7516/2021/10/043}, \href {https://ui.adsabs.harvard.edu/abs/2021JCAP...10..043B} {2021, 043}

\bibitem[\protect\citeauthoryear{{Bode}, {Ostriker}  \& {Turok}}{{Bode} et~al.}{2001}]{Bode_etal_2001}
{Bode} P.,  {Ostriker} J.~P.,   {Turok} N.,  2001, \mn@doi [\apj] {10.1086/321541}, \href {https://ui.adsabs.harvard.edu/abs/2001ApJ...556...93B} {556, 93}

\bibitem[\protect\citeauthoryear{{Bonaca}, {Hogg}, {Price-Whelan}  \& {Conroy}}{{Bonaca} et~al.}{2019}]{bonaca19}
{Bonaca} A.,  {Hogg} D.~W.,  {Price-Whelan} A.~M.,   {Conroy} C.,  2019, \mn@doi [\apj] {10.3847/1538-4357/ab2873}, \href {https://ui.adsabs.harvard.edu/abs/2019ApJ...880...38B} {880, 38}

\bibitem[\protect\citeauthoryear{{Bond} \& {Szalay}}{{Bond} \& {Szalay}}{1983}]{Bond_Szalay_1983}
{Bond} J.~R.,  {Szalay} A.~S.,  1983, \mn@doi [\apj] {10.1086/161460}, \href {https://ui.adsabs.harvard.edu/abs/1983ApJ...274..443B} {274, 443}

\bibitem[\protect\citeauthoryear{{Bovy}, {Erkal}  \& {Sanders}}{{Bovy} et~al.}{2017}]{Bovy_etal_2017}
{Bovy} J.,  {Erkal} D.,   {Sanders} J.~L.,  2017, \mn@doi [\mnras] {10.1093/mnras/stw3067}, \href {https://ui.adsabs.harvard.edu/abs/2017MNRAS.466..628B} {466, 628}

\bibitem[\protect\citeauthoryear{{Carlberg}}{{Carlberg}}{2009}]{carlberg09}
{Carlberg} R.~G.,  2009, \mn@doi [\apjl] {10.1088/0004-637X/705/2/L223}, \href {https://ui.adsabs.harvard.edu/abs/2009ApJ...705L.223C} {705, L223}

\bibitem[\protect\citeauthoryear{{Carlberg}}{{Carlberg}}{2012}]{Carlberg_2012}
{Carlberg} R.~G.,  2012, \mn@doi [\apj] {10.1088/0004-637X/748/1/20}, \href {https://ui.adsabs.harvard.edu/abs/2012ApJ...748...20C} {748, 20}

\bibitem[\protect\citeauthoryear{{Chabrier}}{{Chabrier}}{2005}]{chabrier05}
{Chabrier} G.,  2005, in {Corbelli} E.,  {Palla} F.,   {Zinnecker} H.,  eds,  Astrophysics and Space Science Library Vol. 327, The Initial Mass Function 50 Years Later. p.~41 (\mn@eprint {arXiv} {astro-ph/0409465}), \mn@doi{10.1007/978-1-4020-3407-7_5}

\bibitem[\protect\citeauthoryear{{Choi}, {Dotter}, {Conroy}, {Cantiello}, {Paxton}  \& {Johnson}}{{Choi} et~al.}{2016}]{choi16}
{Choi} J.,  {Dotter} A.,  {Conroy} C.,  {Cantiello} M.,  {Paxton} B.,   {Johnson} B.~D.,  2016, \mn@doi [\apj] {10.3847/0004-637X/823/2/102}, \href {https://ui.adsabs.harvard.edu/abs/2016ApJ...823..102C} {823, 102}

\bibitem[\protect\citeauthoryear{{Dehnen} \& {Binney}}{{Dehnen} \& {Binney}}{1998}]{dehnen98b}
{Dehnen} W.,  {Binney} J.,  1998, \mn@doi [\mnras] {10.1046/j.1365-8711.1998.01282.x}, \href {https://ui.adsabs.harvard.edu/abs/1998MNRAS.294..429D} {294, 429}

\bibitem[\protect\citeauthoryear{{Dotter}}{{Dotter}}{2016}]{dotter16}
{Dotter} A.,  2016, \mn@doi [\apjs] {10.3847/0067-0049/222/1/8}, \href {https://ui.adsabs.harvard.edu/abs/2016ApJS..222....8D} {222, 8}

\bibitem[\protect\citeauthoryear{{Erkal} \& {Belokurov}}{{Erkal} \& {Belokurov}}{2015a}]{erkal15a}
{Erkal} D.,  {Belokurov} V.,  2015a, \mn@doi [\mnras] {10.1093/mnras/stv655}, \href {https://ui.adsabs.harvard.edu/abs/2015MNRAS.450.1136E} {450, 1136}

\bibitem[\protect\citeauthoryear{{Erkal} \& {Belokurov}}{{Erkal} \& {Belokurov}}{2015b}]{erkal15b}
{Erkal} D.,  {Belokurov} V.,  2015b, \mn@doi [\mnras] {10.1093/mnras/stv2122}, \href {https://ui.adsabs.harvard.edu/abs/2015MNRAS.454.3542E} {454, 3542}

\bibitem[\protect\citeauthoryear{{Erkal} \& {Belokurov}}{{Erkal} \& {Belokurov}}{2020}]{erkal2020}
{Erkal} D.,  {Belokurov} V.~A.,  2020, \mn@doi [\mnras] {10.1093/mnras/staa1238}, \href {https://ui.adsabs.harvard.edu/abs/2020MNRAS.495.2554E} {495, 2554}

\bibitem[\protect\citeauthoryear{{Erkal}, {Belokurov}, {Bovy}  \& {Sanders}}{{Erkal} et~al.}{2016}]{erkal16}
{Erkal} D.,  {Belokurov} V.,  {Bovy} J.,   {Sanders} J.~L.,  2016, \mn@doi [\mnras] {10.1093/mnras/stw1957}, \href {https://ui.adsabs.harvard.edu/abs/2016MNRAS.463..102E} {463, 102}

\bibitem[\protect\citeauthoryear{{Erkal}, {Koposov}  \& {Belokurov}}{{Erkal} et~al.}{2017}]{erkal2017}
{Erkal} D.,  {Koposov} S.~E.,   {Belokurov} V.,  2017, \mn@doi [\mnras] {10.1093/mnras/stx1208}, \href {https://ui.adsabs.harvard.edu/abs/2017MNRAS.470...60E} {470, 60}

\bibitem[\protect\citeauthoryear{{Erkal} et~al.,}{{Erkal} et~al.}{2019}]{erkal19}
{Erkal} D.,  et~al., 2019, \mn@doi [\mnras] {10.1093/mnras/stz1371}, \href {https://ui.adsabs.harvard.edu/abs/2019MNRAS.487.2685E} {487, 2685}

\bibitem[\protect\citeauthoryear{{Erkal} et~al.,}{{Erkal} et~al.}{2021}]{erkal21}
{Erkal} D.,  et~al., 2021, \mn@doi [\mnras] {10.1093/mnras/stab1828}, \href {https://ui.adsabs.harvard.edu/abs/2021MNRAS.506.2677E} {506, 2677}

\bibitem[\protect\citeauthoryear{{Eyer} et~al.,}{{Eyer} et~al.}{2012}]{eyer12}
{Eyer} L.,  et~al., 2012, in {Richards} M.~T.,  {Hubeny} I.,  eds, ~ Vol. 282, From Interacting Binaries to Exoplanets: Essential Modeling Tools. pp 33--40 (\mn@eprint {arXiv} {1201.5140}), \mn@doi{10.1017/S1743921311026822}

\bibitem[\protect\citeauthoryear{{Foreman-Mackey}, {Hogg}, {Lang}  \& {Goodman}}{{Foreman-Mackey} et~al.}{2013}]{2013PASP..125..306F}
{Foreman-Mackey} D.,  {Hogg} D.~W.,  {Lang} D.,   {Goodman} J.,  2013, \mn@doi [\pasp] {10.1086/670067}, \href {https://ui.adsabs.harvard.edu/abs/2013PASP..125..306F} {125, 306}

\bibitem[\protect\citeauthoryear{{Gaia Collaboration} et~al.,}{{Gaia Collaboration} et~al.}{2023}]{gaia23}
{Gaia Collaboration} et~al., 2023, \mn@doi [\aap] {10.1051/0004-6361/202243940}, \href {https://ui.adsabs.harvard.edu/abs/2023A&A...674A...1G} {674, A1}

\bibitem[\protect\citeauthoryear{{Gibbons}, {Belokurov}  \& {Evans}}{{Gibbons} et~al.}{2014}]{gibbons14}
{Gibbons} S.~L.~J.,  {Belokurov} V.,   {Evans} N.~W.,  2014, \mn@doi [\mnras] {10.1093/mnras/stu1986}, \href {https://ui.adsabs.harvard.edu/abs/2014MNRAS.445.3788G} {445, 3788}

\bibitem[\protect\citeauthoryear{{Gieles}, {Erkal}, {Antonini}, {Balbinot}  \& {Pe{\~n}arrubia}}{{Gieles} et~al.}{2021}]{Gieles+2021}
{Gieles} M.,  {Erkal} D.,  {Antonini} F.,  {Balbinot} E.,   {Pe{\~n}arrubia} J.,  2021, \mn@doi [Nature Astronomy] {10.1038/s41550-021-01392-2}, \href {https://ui.adsabs.harvard.edu/abs/2021NatAs...5..957G} {5, 957}

\bibitem[\protect\citeauthoryear{{G{\'o}mez}, {Besla}, {Carpintero}, {Villalobos}, {O'Shea}  \& {Bell}}{{G{\'o}mez} et~al.}{2015}]{gomez15}
{G{\'o}mez} F.~A.,  {Besla} G.,  {Carpintero} D.~D.,  {Villalobos} {\'A}.,  {O'Shea} B.~W.,   {Bell} E.~F.,  2015, \mn@doi [\apj] {10.1088/0004-637X/802/2/128}, \href {https://ui.adsabs.harvard.edu/abs/2015ApJ...802..128G} {802, 128}

\bibitem[\protect\citeauthoryear{{Grillmair} \& {Dionatos}}{{Grillmair} \& {Dionatos}}{2006}]{GD1_disc}
{Grillmair} C.~J.,  {Dionatos} O.,  2006, \mn@doi [\apjl] {10.1086/505111}, \href {https://ui.adsabs.harvard.edu/abs/2006ApJ...643L..17G} {643, L17}

\bibitem[\protect\citeauthoryear{{Hernquist}}{{Hernquist}}{1990}]{hernquist90}
{Hernquist} L.,  1990, \mn@doi [\apj] {10.1086/168845}, \href {https://ui.adsabs.harvard.edu/abs/1990ApJ...356..359H} {356, 359}

\bibitem[\protect\citeauthoryear{{Hui}, {Ostriker}, {Tremaine}  \& {Witten}}{{Hui} et~al.}{2017}]{Hui_etal_2017}
{Hui} L.,  {Ostriker} J.~P.,  {Tremaine} S.,   {Witten} E.,  2017, \mn@doi [\prd] {10.1103/PhysRevD.95.043541}, \href {https://ui.adsabs.harvard.edu/abs/2017PhRvD..95d3541H} {95, 043541}

\bibitem[\protect\citeauthoryear{{Ibata}, {Lewis}, {Irwin}  \& {Quinn}}{{Ibata} et~al.}{2002}]{ibata02}
{Ibata} R.~A.,  {Lewis} G.~F.,  {Irwin} M.~J.,   {Quinn} T.,  2002, \mn@doi [\mnras] {10.1046/j.1365-8711.2002.05358.x}, \href {https://ui.adsabs.harvard.edu/abs/2002MNRAS.332..915I} {332, 915}

\bibitem[\protect\citeauthoryear{{Jethwa}, {Erkal}  \& {Belokurov}}{{Jethwa} et~al.}{2018}]{Jethwa_etal_2018}
{Jethwa} P.,  {Erkal} D.,   {Belokurov} V.,  2018, \mn@doi [\mnras] {10.1093/mnras/stx2330}, \href {https://ui.adsabs.harvard.edu/abs/2018MNRAS.473.2060J} {473, 2060}

\bibitem[\protect\citeauthoryear{{Johnston}, {Spergel}  \& {Haydn}}{{Johnston} et~al.}{2002}]{johnston02}
{Johnston} K.~V.,  {Spergel} D.~N.,   {Haydn} C.,  2002, \mn@doi [\apj] {10.1086/339791}, \href {https://ui.adsabs.harvard.edu/abs/2002ApJ...570..656J} {570, 656}

\bibitem[\protect\citeauthoryear{{Klypin}, {Kravtsov}, {Valenzuela}  \& {Prada}}{{Klypin} et~al.}{1999}]{klypin1999}
{Klypin} A.,  {Kravtsov} A.~V.,  {Valenzuela} O.,   {Prada} F.,  1999, \mn@doi [\apj] {10.1086/307643}, \href {https://ui.adsabs.harvard.edu/abs/1999ApJ...522...82K} {522, 82}

\bibitem[\protect\citeauthoryear{Koposov}{Koposov}{2021}]{koposovsoftware}
Koposov S.,  2021, segasai/minimint: Minimint 0.3.0, \mn@doi{10.5281/zenodo.5610692}, \url {https://doi.org/10.5281/zenodo.5610692}

\bibitem[\protect\citeauthoryear{{Koposov}, {Irwin}, {Belokurov}, {Gonzalez-Solares}, {Yoldas}, {Lewis}, {Metcalfe}  \& {Shanks}}{{Koposov} et~al.}{2014}]{koposov14}
{Koposov} S.~E.,  {Irwin} M.,  {Belokurov} V.,  {Gonzalez-Solares} E.,  {Yoldas} A.~K.,  {Lewis} J.,  {Metcalfe} N.,   {Shanks} T.,  2014, \mn@doi [\mnras] {10.1093/mnrasl/slu060}, \href {https://ui.adsabs.harvard.edu/abs/2014MNRAS.442L..85K} {442, L85}

\bibitem[\protect\citeauthoryear{{Koposov} et~al.,}{{Koposov} et~al.}{2023}]{koposov23}
{Koposov} S.~E.,  et~al., 2023, \mn@doi [\mnras] {10.1093/mnras/stad551}, \href {https://ui.adsabs.harvard.edu/abs/2023MNRAS.521.4936K} {521, 4936}

\bibitem[\protect\citeauthoryear{{Li} \& {S5 Collaboration}}{{Li} \& {S5 Collaboration}}{2021}]{S5_DR1}
{Li} T.,  {S5 Collaboration} 2021, {Southern Stellar Stream Spectroscopic Survey: The First Public Data Release}, \mn@doi{10.5281/zenodo.4695135}, \url {https://doi.org/10.5281/zenodo.4695135}

\bibitem[\protect\citeauthoryear{{Li} et~al.,}{{Li} et~al.}{2019}]{li19}
{Li} T.~S.,  et~al., 2019, \mn@doi [\mnras] {10.1093/mnras/stz2731}, \href {https://ui.adsabs.harvard.edu/abs/2019MNRAS.490.3508L} {490, 3508}

\bibitem[\protect\citeauthoryear{{Li} et~al.,}{{Li} et~al.}{2021}]{li21}
{Li} T.~S.,  et~al., 2021, \mn@doi [\apj] {10.3847/1538-4357/abeb18}, \href {https://ui.adsabs.harvard.edu/abs/2021ApJ...911..149L} {911, 149}

\bibitem[\protect\citeauthoryear{{Li} et~al.,}{{Li} et~al.}{2022}]{li22}
{Li} T.~S.,  et~al., 2022, \mn@doi [\apj] {10.3847/1538-4357/ac46d3}, \href {https://ui.adsabs.harvard.edu/abs/2022ApJ...928...30L} {928, 30}

\bibitem[\protect\citeauthoryear{{Malhan}, {Valluri}  \& {Freese}}{{Malhan} et~al.}{2020}]{malhan20}
{Malhan} K.,  {Valluri} M.,   {Freese} K.,  2020, arXiv e-prints, \href {https://ui.adsabs.harvard.edu/abs/2020arXiv200512919M} {p. arXiv:2005.12919}

\bibitem[\protect\citeauthoryear{{McMillan}}{{McMillan}}{2017}]{mcmillan17}
{McMillan} P.~J.,  2017, \mn@doi [\mnras] {10.1093/mnras/stw2759}, \href {https://ui.adsabs.harvard.edu/abs/2017MNRAS.465...76M} {465, 76}

\bibitem[\protect\citeauthoryear{{Mirabal} \& {Bonaca}}{{Mirabal} \& {Bonaca}}{2021}]{Mirabel_Bonaca_2021}
{Mirabal} N.,  {Bonaca} A.,  2021, \mn@doi [\jcap] {10.1088/1475-7516/2021/11/033}, \href {https://ui.adsabs.harvard.edu/abs/2021JCAP...11..033M} {2021, 033}

\bibitem[\protect\citeauthoryear{{Moore}, {Ghigna}, {Governato}, {Lake}, {Quinn}, {Stadel}  \& {Tozzi}}{{Moore} et~al.}{1999}]{moore99}
{Moore} B.,  {Ghigna} S.,  {Governato} F.,  {Lake} G.,  {Quinn} T.,  {Stadel} J.,   {Tozzi} P.,  1999, \mn@doi [\apjl] {10.1086/312287}, \href {https://ui.adsabs.harvard.edu/abs/1999ApJ...524L..19M} {524, L19}

\bibitem[\protect\citeauthoryear{{Nadler} et~al.,}{{Nadler} et~al.}{2021}]{Nadler_etal_2021}
{Nadler} E.~O.,  et~al., 2021, \mn@doi [\prl] {10.1103/PhysRevLett.126.091101}, \href {https://ui.adsabs.harvard.edu/abs/2021PhRvL.126i1101N} {126, 091101}

\bibitem[\protect\citeauthoryear{{Navarro}, {Frenk}  \& {White}}{{Navarro} et~al.}{1997}]{navarro97}
{Navarro} J.~F.,  {Frenk} C.~S.,   {White} S. D.~M.,  1997, \mn@doi [\apj] {10.1086/304888}, \href {https://ui.adsabs.harvard.edu/abs/1997ApJ...490..493N} {490, 493}

\bibitem[\protect\citeauthoryear{{Odenkirchen} et~al.,}{{Odenkirchen} et~al.}{2001}]{Pal5_disc}
{Odenkirchen} M.,  et~al., 2001, \mn@doi [\apjl] {10.1086/319095}, \href {https://ui.adsabs.harvard.edu/abs/2001ApJ...548L.165O} {548, L165}

\bibitem[\protect\citeauthoryear{{Plummer}}{{Plummer}}{1911}]{Plummer_1911}
{Plummer} H.~C.,  1911, \mn@doi [\mnras] {10.1093/mnras/71.5.460}, \href {https://ui.adsabs.harvard.edu/abs/1911MNRAS..71..460P} {71, 460}

\bibitem[\protect\citeauthoryear{{Price-Whelan} \& {Bonaca}}{{Price-Whelan} \& {Bonaca}}{2018}]{pricewhelan18}
{Price-Whelan} A.~M.,  {Bonaca} A.,  2018, \mn@doi [\apjl] {10.3847/2041-8213/aad7b5}, \href {https://ui.adsabs.harvard.edu/abs/2018ApJ...863L..20P} {863, L20}

\bibitem[\protect\citeauthoryear{{Shanks} et~al.,}{{Shanks} et~al.}{2015}]{Shanks_etal_2015}
{Shanks} T.,  et~al., 2015, \mn@doi [\mnras] {10.1093/mnras/stv1130}, \href {https://ui.adsabs.harvard.edu/abs/2015MNRAS.451.4238S} {451, 4238}

\bibitem[\protect\citeauthoryear{{Shipp} et~al.,}{{Shipp} et~al.}{2018}]{shipp18}
{Shipp} N.,  et~al., 2018, \mn@doi [\apj] {10.3847/1538-4357/aacdab}, \href {https://ui.adsabs.harvard.edu/abs/2018ApJ...862..114S} {862, 114}

\bibitem[\protect\citeauthoryear{{Shipp} et~al.,}{{Shipp} et~al.}{2019}]{shipp19}
{Shipp} N.,  et~al., 2019, \mn@doi [\apj] {10.3847/1538-4357/ab44bf}, \href {https://ui.adsabs.harvard.edu/abs/2019ApJ...885....3S} {885, 3}

\bibitem[\protect\citeauthoryear{{Shipp} et~al.,}{{Shipp} et~al.}{2021}]{shipp21}
{Shipp} N.,  et~al., 2021, \mn@doi [\apj] {10.3847/1538-4357/ac2e93}, \href {https://ui.adsabs.harvard.edu/abs/2021ApJ...923..149S} {923, 149}

\bibitem[\protect\citeauthoryear{{Siegal-Gaskins} \& {Valluri}}{{Siegal-Gaskins} \& {Valluri}}{2008}]{siegal08}
{Siegal-Gaskins} J.~M.,  {Valluri} M.,  2008, \mn@doi [\apj] {10.1086/587450}, \href {https://ui.adsabs.harvard.edu/abs/2008ApJ...681...40S} {681, 40}

\bibitem[\protect\citeauthoryear{{Somerville} \& {Dav{\'e}}}{{Somerville} \& {Dav{\'e}}}{2015}]{Somerville_etal_2015}
{Somerville} R.~S.,  {Dav{\'e}} R.,  2015, \mn@doi [\araa] {10.1146/annurev-astro-082812-140951}, \href {https://ui.adsabs.harvard.edu/abs/2015ARA&A..53...51S} {53, 51}

\bibitem[\protect\citeauthoryear{{Springel} et~al.,}{{Springel} et~al.}{2008}]{springel08}
{Springel} V.,  et~al., 2008, \mn@doi [\mnras] {10.1111/j.1365-2966.2008.14066.x}, \href {https://ui.adsabs.harvard.edu/abs/2008MNRAS.391.1685S} {391, 1685}

\bibitem[\protect\citeauthoryear{{Vasiliev}, {Belokurov}  \& {Erkal}}{{Vasiliev} et~al.}{2021}]{vasiliev21}
{Vasiliev} E.,  {Belokurov} V.,   {Erkal} D.,  2021, \mn@doi [\mnras] {10.1093/mnras/staa3673}, \href {https://ui.adsabs.harvard.edu/abs/2021MNRAS.501.2279V} {501, 2279}

\bibitem[\protect\citeauthoryear{{Watkins}, {van der Marel}  \& {Bennet}}{{Watkins} et~al.}{2024}]{watkins24}
{Watkins} L.~L.,  {van der Marel} R.~P.,   {Bennet} P.,  2024, \mn@doi [\apj] {10.3847/1538-4357/ad1f58}, \href {https://ui.adsabs.harvard.edu/abs/2024ApJ...963...84W} {963, 84}

\bibitem[\protect\citeauthoryear{{White} \& {Rees}}{{White} \& {Rees}}{1978}]{1978MNRAS.183..341W}
{White} S.~D.~M.,  {Rees} M.~J.,  1978, \mn@doi [\mnras] {10.1093/mnras/183.3.341}, \href {https://ui.adsabs.harvard.edu/abs/1978MNRAS.183..341W} {183, 341}

\bibitem[\protect\citeauthoryear{{Yoon}, {Johnston}  \& {Hogg}}{{Yoon} et~al.}{2011}]{yoon11}
{Yoon} J.~H.,  {Johnston} K.~V.,   {Hogg} D.~W.,  2011, \mn@doi [\apj] {10.1088/0004-637X/731/1/58}, \href {https://ui.adsabs.harvard.edu/abs/2011ApJ...731...58Y} {731, 58}

\bibitem[\protect\citeauthoryear{{de Jong} et~al.,}{{de Jong} et~al.}{2012}]{dejong12}
{de Jong} R.~S.,  et~al., 2012, in {McLean} I.~S.,  {Ramsay} S.~K.,   {Takami} H.,  eds,  Society of Photo-Optical Instrumentation Engineers (SPIE) Conference Series Vol. 8446, Ground-based and Airborne Instrumentation for Astronomy IV. p. 84460T (\mn@eprint {arXiv} {1206.6885}), \mn@doi{10.1117/12.926239}

\bibitem[\protect\citeauthoryear{{van der Marel} \& {Kallivayalil}}{{van der Marel} \& {Kallivayalil}}{2014}]{marel14}
{van der Marel} R.~P.,  {Kallivayalil} N.,  2014, \mn@doi [\apj] {10.1088/0004-637X/781/2/121}, \href {https://ui.adsabs.harvard.edu/abs/2014ApJ...781..121V} {781, 121}

\makeatother
\end{thebibliography}




\appendix

\section{Mock observables}

In this appendix, we show mock observations of our simulated stream in three different observational setups. Figure~\ref{fig:presentobservables} shows mock observations with no additional observational uncertainties, i.e. the uncertainties come from the Poisson uncertainties associated with the finite number of particles in our mock stream. Figure~\ref{fig:presentobservables} shows the mock observations of our simulated stream assuming present-day errors and Figure~\ref{fig:futureobservables} shows the mock observations of our simulated stream assuming future observational errors, i.e. with the proper motion uncertainties assuming 10 years of data from \textit{Gaia} and with the expected radial velocity uncertainty and depth from 4MOST.

\begin{figure}
 \includegraphics[width=\columnwidth]{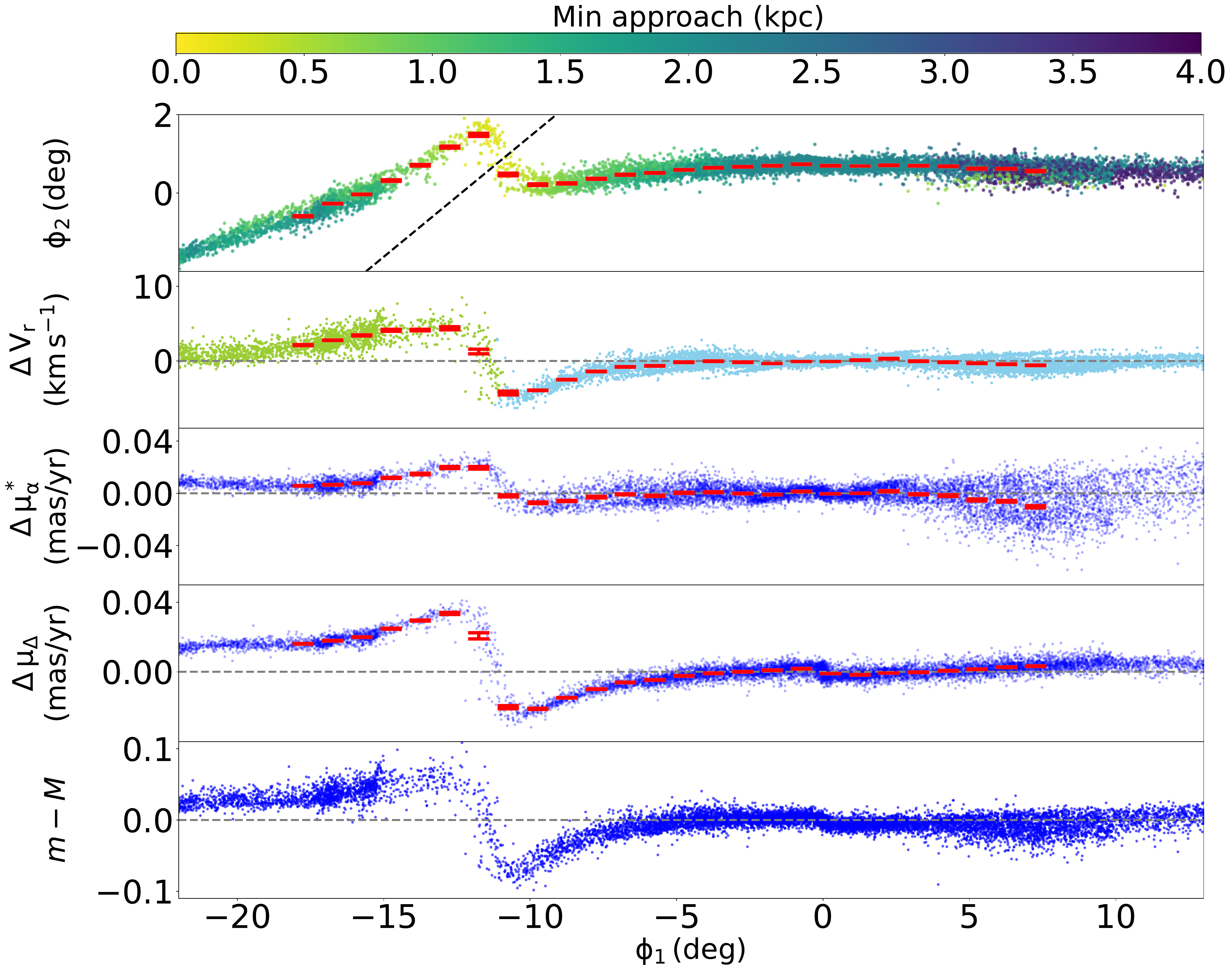}
 \caption{The data of our mock perturbed stream with no observational errors, errorbars instead showing the spread in $\mathrm{1\,\deg}$ bins in $\mathrm{\phi_1}$. An exception is made in the stream track, as explained in Section \ref{sec:mocksummary} where we inflate the error to account for minor variations in the random seed. Top: Stream track, with the minimum distance achieved between individual mock stream stars and the subhalo shown in the colour bar. The black data points highlight the radial velocity and proper motions of stars in the panels below. We also show a dotted line that separates the observed stars as belonging to either the ATLAS half of the stream (right of the kink) or the Aliqa Uma half (left of the kink). Second row: The radial velocity differences between the mock perturbed stream and a fourth-order polynomial fit of $\mathrm{V_{GSR}}$ for the unperturbed model, with mock observables shown as red error bars. The stars belonging to the ATLAS part of the stream are shown in light blue and Aliqa Uma in green. Third and fourth row: The proper motion differences between the perturbed stream and a fourth-order polynomial fit, with mock observables shown once again as red error bars, much in the same vein as the radial velocities. Bottom row: Distance modulus difference between the perturbed stream and a fifth-order polynomial fit to the unperturbed model.}
 \label{fig:noobservables}
\end{figure}

\begin{figure}
 \includegraphics[width=\columnwidth]{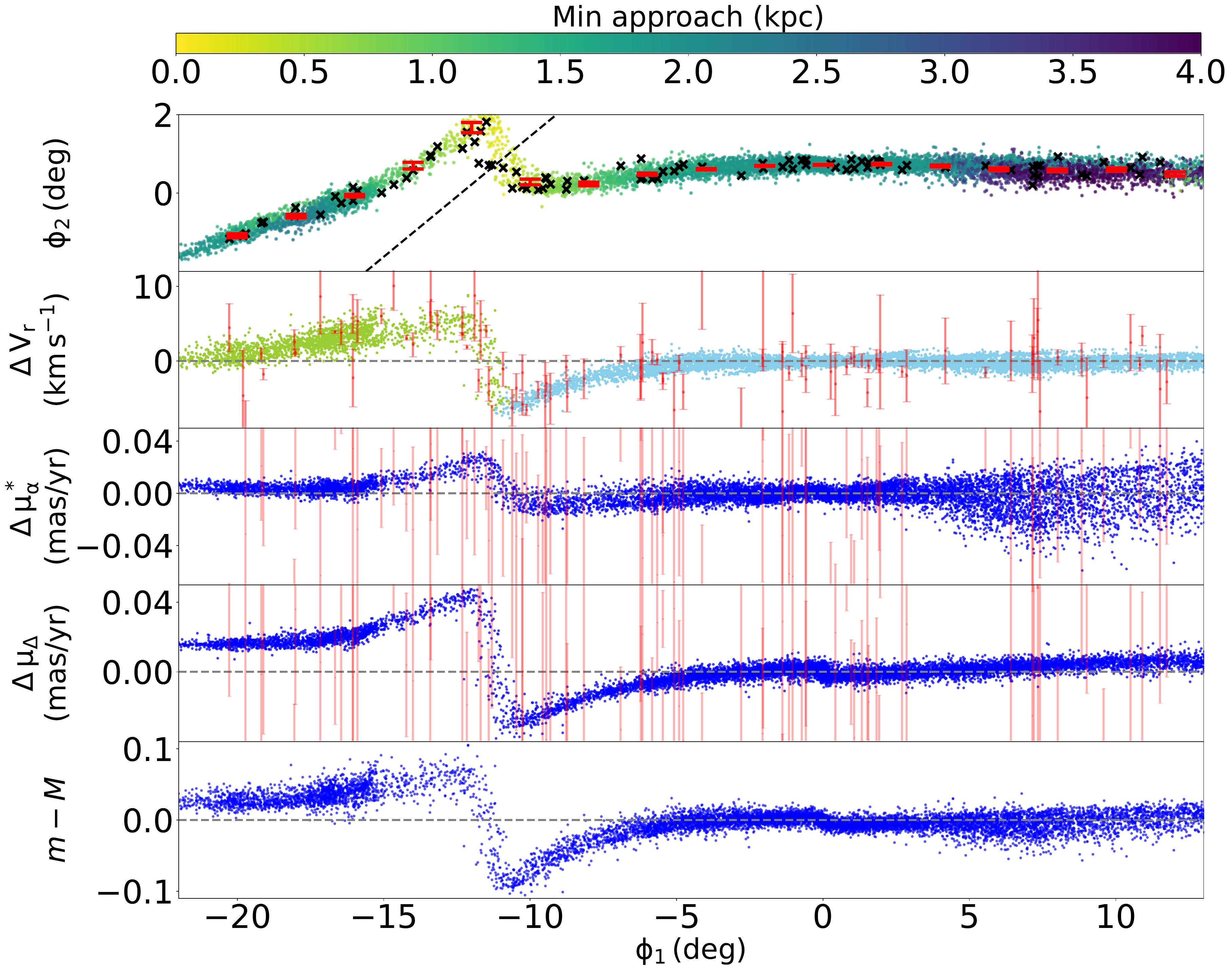}
 \caption{The data of our mock perturbed stream with data following present-day errors overlaid. Top: Stream track, with the minimum distance achieved between individual mock stream stars and the subhalo shown in the colour bar. The black data points highlight the radial velocity and proper motions of stars in the panels below. We also show a dotted line that separates the observed stars as belonging to either the ATLAS half of the stream (right of the kink) or the Aliqa Uma half (left of the kink). Second row: The radial velocity differences between the mock perturbed stream and a fourth-order polynomial fit of $\mathrm{V_{GSR}}$ for the unperturbed model, with mock observables shown as red error bars. The stars belonging to the ATLAS part of the stream are shown in light blue, and Aliqa Uma in green. Third and fourth row: The proper motion differences between the perturbed stream and a fourth-order polynomial fit, with mock observables shown once again as red error bars, much in the same vein as the radial velocities. Bottom row: Distance modulus difference between the perturbed stream and a fifth-order polynomial fit to the unperturbed model.}
 \label{fig:presentobservables}
\end{figure}

\begin{figure}
 \includegraphics[width=\columnwidth]{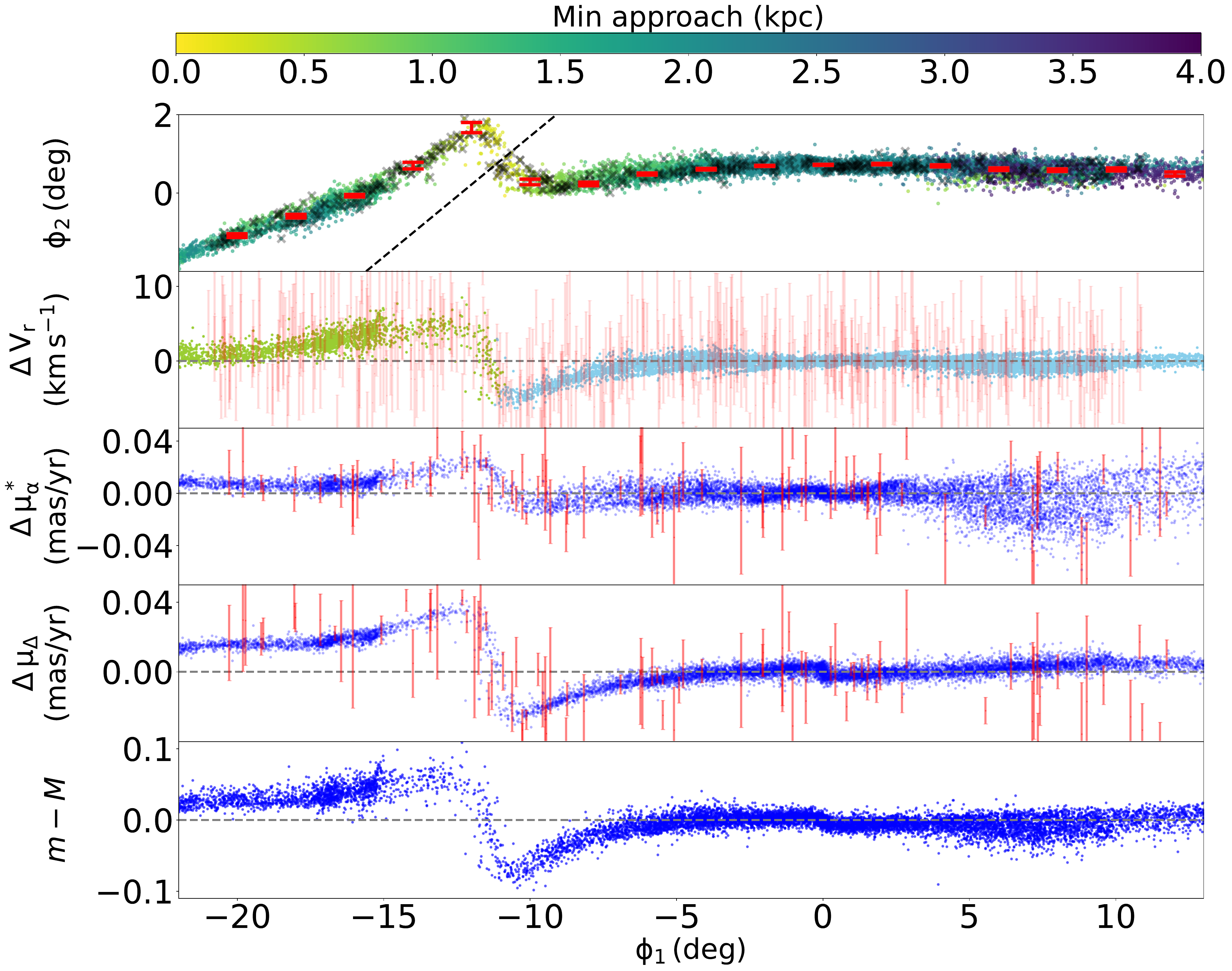}
 \caption{The data of our mock perturbed stream with data following future predicted errors overlaid. Top: Stream track, with the minimum distance achieved between individual mock stream stars and the subhalo shown in the colour bar. The black data points highlight the radial velocity and proper motions of stars in the panels below. We also show a dotted line that separates the observed stars as belonging to either the ATLAS half of the stream (right of the kink) or the Aliqa Uma half (left of the kink). Second row: The radial velocity differences between the mock perturbed stream and a fourth-order polynomial fit of $\mathrm{V_{GSR}}$ for the unperturbed model, with mock observables shown as red error bars. The stars belonging to the ATLAS part of the stream are shown in light blue and Aliqa Uma in green. Third and fourth row: The proper motion differences between the perturbed stream and a fourth-order polynomial fit, with mock observables shown once again as red error bars, much in the same vein as the radial velocities. Bottom row: Distance modulus difference between the perturbed stream and a fifth-order polynomial fit to the unperturbed model.}
 \label{fig:futureobservables}
\end{figure}

\begin{figure}
 \includegraphics[width=\columnwidth]{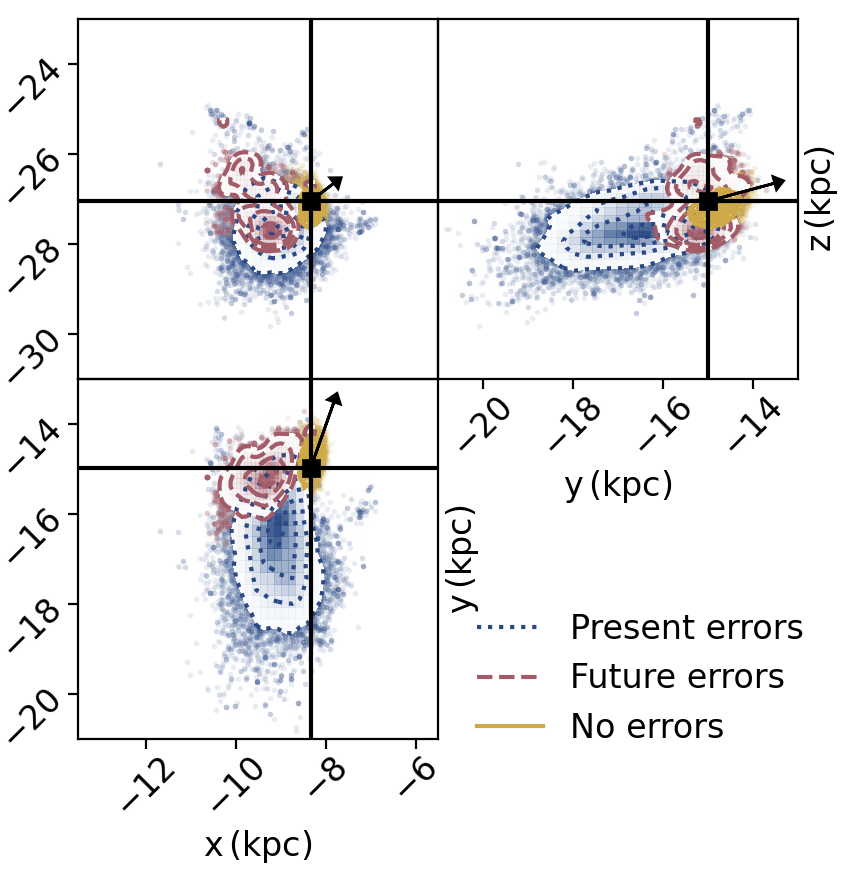}
 \caption{The recovery of the subhalo in the present day for 2500 randomly sampled MCMC runs for all 3 scenarios, with the positions shown in galactocentric coordinates. Yellow: No observational errors, Blue: Present-day observational errors, Red: Future predicted observational errors. The contour levels represent the sigma levels for all 2500 steps, and the scatter shows the full range of possible subhalo positions from the subhalo-stream encounters used. The subhalo position of the mock impact is shown in black, with its direction of orbit also shown.}
 \label{fig:subhalopositionsXYZ}
\end{figure}


\bsp	
\label{lastpage}
\end{document}